\RequirePackage{fix-cm}
\documentclass[smallextended]{svjour3}       
\smartqed  
\usepackage{graphicx}
\usepackage{framed}
\usepackage[justification=centering]{caption}
\usepackage{color}
\usepackage{multirow}
\usepackage{url}
\begin{document}

\title{Beyond the Virus: A First Look at Coronavirus-themed Android Malware}

\author{Liu Wang\and
        Ren He\and
        Haoyu Wang\and
        Pengcheng Xia\and
        Yuanchun Li\and
        Lei Wu\and
        Yajin Zhou\and
        Xiapu Luo\and
        Yulei Sui\and
        Yao Guo\and
        Guoai Xu
}

\institute{Liu Wang, Ren He, Haoyu Wang, Pengcheng Xia, Guoai Xu\at
              Beijing University of Posts and Telecommunications, Beijing, China\\
              \email{haoyuwang@bupt.edu.cn} \\ \\
              Yuanchun Li, Microsoft Research Asia, Beijing, China \\
              Lei Wu, Yajin Zhou, Zhejiang University, Hangzhou, China\\
              Xiapu Luo, The Hong Kong Polytechnic University, HongKong, China\\
              Yulei Sui, University of Technology Sydney, Australia\\
              Yao Guo, Peking University, Beijing, China\\
}

\date{Received: date / Accepted: date}

\maketitle

\begin{abstract}

As the COVID-19 pandemic emerged in early 2020, a number of malicious actors have started capitalizing the topic. Although a few media reports mentioned the existence of coronavirus-themed mobile malware, the research community lacks the understanding of the landscape of the coronavirus-themed mobile malware. In this paper, we present the first systematic study of coronavirus-themed Android malware. We first make efforts to create a daily growing COVID-19 themed mobile app dataset, which contains $4,322$ COVID-19 themed apk samples (2,500 unique apps) and $611$ potential malware samples (370 unique malicious apps) by the time of mid-November, 2020.
We then present an analysis of them from multiple perspectives including trends and statistics, installation methods, malicious behaviors and malicious actors behind them. 
We observe that the COVID-19 themed apps as well as malicious ones began to flourish almost as soon as the pandemic broke out worldwide.
Most malicious apps are camouflaged as benign apps using the same app identifiers (e.g., app name, package name and app icon).
Their main purposes are either stealing users' private information or making profit by using tricks like phishing and extortion. 
Furthermore, only a quarter of the COVID-19 malware creators are habitual developers who have been active for a long time, while 75\% of them are newcomers in this pandemic. 
The malicious developers are mainly located in the US, mostly targeting countries including English-speaking countries, China, Arabic countries and Europe.
\textit{To facilitate future research, we have publicly released all the well-labelled COVID-19 themed apps (and malware) to the research community. Till now, over 30 research institutes around the world have requested our dataset for COVID-19 themed research.}


\end{abstract}

\section{Introduction}

As COVID-19 continues to spread across the world, a growing number of malicious actors are exploiting the pandemic to make a profit. It is reported that COVID-19 is being used in a variety of online malicious activities, including Email scam, donation scam, ransomware and phishing websites~\cite{trendmicronews,paloalto,NCSC,bankinfosecurity,fireeye}. As the number of afflicted cases continue to surge, malicious activities that use coronavirus as a lure are increasing. 

There is little doubt that smartphones have become a central part of our lives, allowing us to perform all sorts of tasks that make our everyday existence easier and more enjoyable, but they are also becoming a bigger target for cybercriminals~\cite{MYSTORY}. At this particular time of the crisis, smartphones are one of the most popular ways for people to keep track of the most up-to-date status of the pandemic, receive notifications, learn about actions for avoiding infections, etc~\cite{scienceApp,IYENGAR2020733}.
Thus, malicious developers take advantage of this opportunity to lure mobile users to download and install malware and potentially harmful apps (PHAs). Indeed, some existing news reports~\cite{androidreport1,androidreport2,androidreport3,androidreport4} show that COVID-19 related malicious apps have been observed, and thousands of mobile users have been affected in another way (by the virtual virus) in this pandemic. 
For example, the malicious website (\texttt{coronavirusapp.site}) prompts users to download a malicious Android app that will give them access to a coronavirus map tracker that appears to provide tracking and statistical information about COVID-19. However, the app is indeed a ransomware that locks users' screen, which requests \$100 in Bitcoin to unlock the phone.

However, besides a few media reports, the coronavirus-themed mobile malware have not been well studied by the research community. Our community lacks the comprehensive understanding of the landscape of the coronavirus-themed mobile malware, and no accessible dataset could be used by researchers to boost COVID-19 related cybersecurity studies. 

\textbf{This Work.}
To this end, this paper presents the first measurement study of COVID-19 related Android malware.
We first make efforts to create a daily growing COVID-19 related mobile app dataset (see \textbf{Section~\ref{sec:dataset}}), by collecting samples from a number of sources, including app markets (both Google Play and alternative app markets), a well-known app repository (i.e., Koodous), the COVID-19 related domains (apps downloaded or connected to these domains), and security threat intelligence reports. 
We have released the dataset to the community since May 29th, 2020. To the best of our knowledge, this is the first COVID-19 themed mobile apps and malware dataset.
By the time of this paper writing, we have curated a dataset of 4,322 COVID-19 themed apk samples (2,500 different apps), and 611 (370 different malicious apps) of them are considered to be malicious. We then present a comprehensive analysis of these apps from the perspectives including \textit{trends and statistics} (see \textbf{Section~\ref{sec:trends}}), \textit{app creation and installation} (see \textbf{Section~\ref{sec:creation}}), \textit{malicious behaviors} (see \textbf{Section~\ref{sec:behavior}}), and \textit{the malicious actors} behind them (see \textbf{Section~\ref{sec:developers}}).

Among many interesting results and observations, the following are most prominent:

\begin{itemize}
    \item \textbf{COVID-19 themed Android apps and malware are prevalent.}
    We have identified over 4,300 COVID-19 themed Android apk samples (2,500 unique apps according to the package names) by mid-November\footnote{The number is growing daily and our dataset will update weekly.}, and most of them were released after March 15, the time when coronavirus became a pandemic. 
    Among them, 611 samples (370 unique apps) are considered to be malicious.
    
    
    \item 
    \textbf{Fake app is the main way to lure users to install malware.}
    Most of the malicious apps (over 49.6\%) are camouflaged as official apps using the same app identifiers (both app name and package name), and a number of them use confusingly similar app icons to mislead users. However, app repackaging is no longer the main way to create COVID-19 themed Android malware, with only about 5.4\% of them being considered to be repackaged from benign apps.

    \item \textbf{Information Stealing, Phishing and Extortion are the major behaviors of COVID-19 themed Android malware.}
    Trojan and Spyware are two main categories for COVID-19 themed malware. Their purposes are either stealing users' private information, or making a profit using tricks like phishing, premium SMS/Phone calls, stealing bank accounts, and extortion. Besides, anti-analysis techniques are used by roughly 52\% of these malicious apps, yet surprisingly a bit lower than benign apps.
    
    \item \textbf{Although some COVID-19 themed malicious apps are created by experienced actors, most of them are created by emerging developers that target this pandemic.}
    A quarter of the COVID-19 themed malware developers are experienced malicious actors who released malware prior to this pandemic, and the remaining three quarters are emerging malicious developers.
    The COVID-19 is used as a lure to attack unsuspecting users.
    We have collected over 228k apps released by these developers (from 2014 to 2020), and found most of them are malicious. 
    Based on the information extracted from the malicious apps, we find that these developers are mainly located in the US, with rest of them are located in India, Turkey, etc. Besides English-speaking countries, the Arabic countries, Europe, and China are also the main targets of them.
\end{itemize}

To boost the research on coronavirus-themed cybersecurity threats, we have released a daily growing dataset to the research community at:

\begin{center}
\url{https://covid19apps.github.io}
\end{center}

\section{Study Design}

\subsection{Research Questions}
Considering that a number of malicious actors are capitalizing the COVID-19 pandemic in the cyber space, it is thus important to understand the scale and impacts of COVID-19 themed malware, which will further help develop methods to protect users from such attacks. However, no existing studies have shown the trends and characteristics of COVID-19 themed Android malware after the global outbreak of the pandemic. Furthermore, we note that the COVID-19 themed Android malicious apps are rarely developed independently, i.e., some malicious actors may release a number of such apps. Thus, analyzing the COVID-19 themed Android malware from the angle of malicious actors behind them would help us gain deeper understanding of the scope and sophistication of these threats. Thus, our study is driven by the following research questions (RQs):
\begin{itemize}
    \item[RQ1] \textit{How many coronavirus-themed apps are there in the world and how many of them are considered to be potentially harmful?} 
    Existing news reports tend to have poor coverage of COVID-19 malicious apps in the wild.
    Considering that the coronavirus pandemic has been emerging since early 2020, it is thus interesting to investigate \textit{when} the COVID-19 themed apps (and malware) are increasingly popular. 
    We will study this trends and statistics in Section~\ref{sec:trends}.
    
    \item[RQ2]
    It is known to us that the COVID-19 themed malicious apps are taking advantage of the pandemic to attract and lure users to install them. However, it is still unknown to us \textit{how these apps are created and how they can get installed into users' smartphones?} \textit{Whether some social engineering based techniques have been adopted?}
    We will answer this RQ in Section~\ref{sec:creation}.
    
    \item[RQ3]
    Considering that the mobile malware has been widely studied, it is important to study the characteristics of the COVID-19 themed malware. 
    \textit{What are the malicious behaviors of them? What are their purposes? Do they apply any anti-analysis techniques to evade detection?}
    Malicious behaviors will be studied in Section~\ref{sec:behavior}.
    
    \item[RQ4] Given a set of COVID-19 themed malware, \textit{how could we identify the malicious actors behind them? Who created these malware? Are they active for a long time? Who are their main targets?} It would help us gain deeper understanding of the scope and sophistication of these threats. We will answer this RQ in Section~\ref{sec:developers}.
\end{itemize}

\subsection{Dataset Collection}
\label{sec:dataset}

\begin{figure}[t]
\centering
\includegraphics[width=0.99\textwidth]{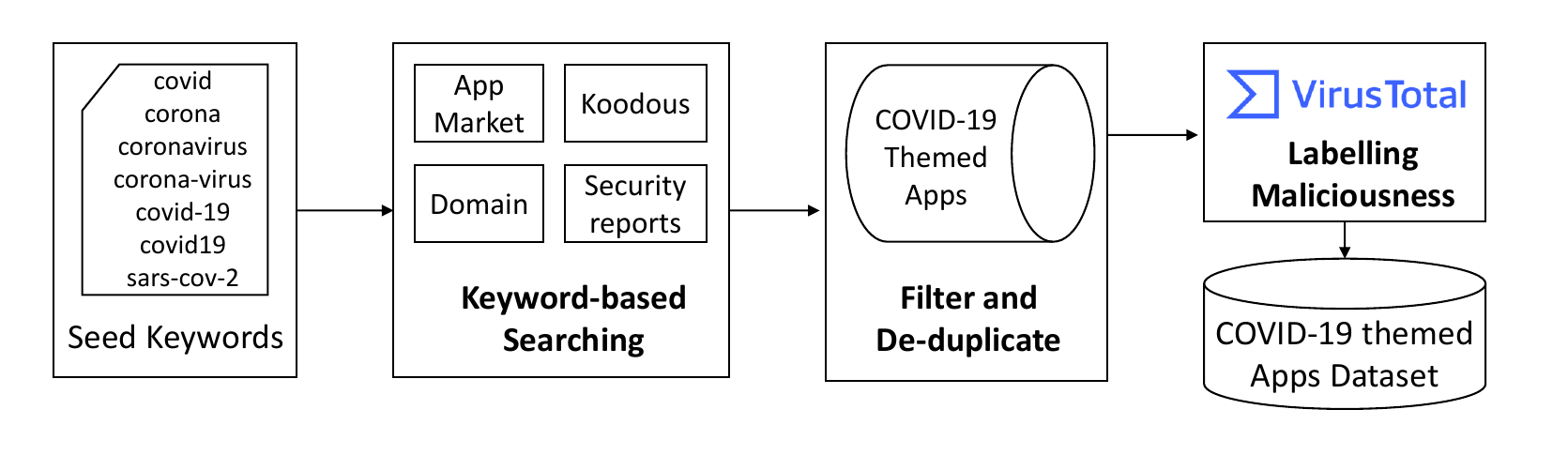}
\caption{Overall process of the dataset collection.}
\label{fig:data-collection} 
\end{figure}

To answer the aforementioned research questions, we first need to harvest a comprehensive dataset of coronavirus-themed apps. 
Considering that a number of malware may be distributed through channels beyond the general app markets~\cite{Farooqi2020UnderstandingIM,MadDroid}, we have adopted a hybrid approach to collect COVID-19 themed apps, and further pinpoint the malicious ones using VirusTotal~\cite{VirusTotal}. Figure~\ref{fig:data-collection} shows the overall process of our dataset collection, which consists of three main steps: keyword-based searching, filtering and de-duplication, and malware labelling. 
We next detail each step.

\subsubsection{Keyword-based Searching} 
Specifically, we collect coronavirus-themed apps from four main sources:
\begin{itemize}
    \item[(1)] \textbf{Android App Markets}. App markets (including Google Play and alternative markets) are general distribution channels for Android mobile apps. Previous research~\cite{wang2018beyond,wang2019understanding} suggested that malicious apps were concurrently found in various app marketplaces.
    Thus, we take into account Google Play~\cite{Googleplay} and 5 alternative markets, including Apkpure~\cite{Apkpure}, Uptodown~\cite{uptodown}, Appchina~\cite{appchina}, Tencent MyApp Market~\cite{MyApp}, and Huawei Market~\cite{Huawei}. Note that Apkpure and Uptodown are two of the most popular alternative app markets for western countries, while Huawei and Tencent MyApp are two of the most popular app markets in China. Thus, we believe our market selection is representative enough. We first manually crafted a set of representative keywords of COVID-19 pandemic, including `covid', `corona', `coronavirus', `corona-virus', `covid-19', `covid19' and `sars-cov-2'. 
    To the best of our knowledge, this set of keywords could cover all the alias of COVID-19, which can be more accurately matched to relevant apps. We admit that other keywords like `facemask' and `lockdown' may be relevant to the COVID-19 themed apps. However, if there are apps containing such keywords but without containing the primary keywords we summarized, we will regard such apps as peripheral apps that are not directly related to COVID-19. We will further discuss the limitation of our keywords selection in Section~\ref{sec:discussion}.
    Then we use these keywords as the seeds to search related apps on Google Play and those 5 alternative markets. For the returned search results we also need to perform a manual check on the app name, package name and app description to identify the truly related apps.
    \item[(2)] \textbf{Existing App Repositories}. 
    Prior work~\cite{Farooqi2020UnderstandingIM,zhou2012dissecting,hu2019dating,hu2020mobile} suggested that app market was not the only source to distribute apps, especially for malware. 
    Malicious apps can be distributed through online forum, Email, SMS, social network, mobile advertisement, and other channels.
    It is non-trivial for us to identify COVID-19 themed malware distributed in these hidden channels. 
    Fortunately, some app repositories provide us a chance to analyze apps beyond app markets.
    To the best of our knowledge, Koodous~\cite{Koodous} is by far the largest Android app repository open to public, with over 69 million apps in total by the end of 2020, and the number is growing rapidly daily. 
    The samples on Koodous are collected from various sources, including app markets, webpages, and thousands of researchers\footnote{Koodous (https://koodous.com/) is designed to be a crowd-sourcing platform for mobile security researchers to share and analyze Android malware.}. Thus, we use the aforementioned keywords to crawl related apps from Koodous, and keep only apps with at least one keyword in the app name or package name. Note that, \textit{although no actual user would download apps from Koodous for installation, we believe the apps from Koodous have value as they were collected in the wild, which can reflect the real-world threats that are distributed through other hidden channels}.
    
    \item[(3)] \textbf{Apps related to the COVID-19 themed domains.} Some apps are distributed through COVID-19 related websites (e.g., \texttt{www.covid19-app.com}). Thus, we take advantage of URLScan~\cite{urlscan}, a URL and website scanner for potentially malicious websites, to collect coronavirus-themed domains first. We use the aforementioned keywords to identify related domains from URLScan, and we have collected 175,966 COVID-19 themed domains.
    Then, we use VirusTotal~\cite{VirusTotal}, an online-service to analyze all the collected domains and get the files related to these domains. 
    For each domain, VirusTotal provides the useful information including files downloaded from this domain, files connected to this domain, and files referred to this domain (i.e., the domain name was hard-coded in the files). We have collected over 1 million related files associated with these domains.
    Note that we only keep the Android apk files whose names or package names contain at least one of our keywords. 
    We further use VirusTotal to collect the metadata information of these apps, e.g., app name, package name, apk file hashing, released date and developer signature, etc.
    
    \item[(4)] \textbf{Security Threat Intelligence Reports.} Threat intelligence platforms (TIPs) are critical security tools that use global security data to help proactively identify, mitigate and remediate security threats~\cite{threat-intelligence-platforms}. TIPs aggregate security intelligence from vendors, analysts and other reputable sources about threats and suspicious activity detected all around the world and share information about viruses, malware and other cyber attacks. Besides, we find that security companies like Mcafee provided reports related to coronavirus themed attacks~\cite{mcafeemalware}. The reports released by TIPs and security companies often contain indicators of compromise (IoCs), which usually provide meta information about malicious apps. Thus, we implement a crawler to fetch all coronavirus-related reports from AlienVault~\cite{AlienVault}, the world’s first truly open threat intelligence community, and extract the meta information of suspicious apps like apk file hash and package names. 
    
\end{itemize}

\subsubsection{Filtering the False Positives.} 
Our keyword-based collection may introduce false positives, e.g., Corona Beer app\footnote{package name: com.corona.extra} would appear in our search results. 
Thus, we further remove the irrelevant apps based on the following two criteria: 
(1) \textit{app release date must be later than December 2019}, as the first confirmed COVID-19 case was in December 2019. Therefore, no coronavirus-themed apps would be released earlier than this time; 
(2) \textit{the apps should not have identical names with well-known brands.}
The official apps released by two famous brands would appear in our search results. The name `Corona' is both the name of a beer brand and a car brand. Thus, we manually remove apps related to this two brands. 

\subsubsection{Labelling the maliciousness of the apps.}
In order to identify the malware, we leverage VirusTotal, a widely-used online malware scanning service aggregating over 60 anti-virus(AV) engines, to scan all the apk files collected. VirusTotal returns a result that shows how many of the AV engines return a positive result (i.e., recognize it as malware), which is called \textit{AV-rank}, and we need to label the scanned apps (malicious or benign) based on it. We then perform a manually random inspection of the apps flagged by only one engine, and find that many of them are truly malicious and valuable for our experiments, so without loss of generality, we choose to treat all apps with AV-rank $\geq 1$ as malware~\cite{piggybacking,ikram2016analysis}.
Following previous measurement studies~\cite{wang2018beyond,ikram2016analysis,wang2019understanding}, we further define the \textit{maliciousness of an app} by the AV-rank, i.e., the larger the AV-rank, the greater the maliciousness.
Then we take advantage of AVClass~\cite{sebastian2016avclass}, a widely used malware labelling tool to get their malware family names (see Section~\ref{sec:behavior}).

\begin{table}[h]
\centering
\small
\caption{Overview of the dataset. The maliciousness of an app is defined by the number of AV engines that recognize it as malware(AV-rank).
}
\resizebox{0.99\textwidth}{!}{
\begin{tabular}{c|c|c|c|c}
\hline\noalign{\smallskip}
Source& \# Apks (apps) &\# Malware (apps)& \# Malicious developers & \# Families\\
\noalign{\smallskip}\hline\noalign{\smallskip}
App Markets & 48(40) & 0 & 0 & 0\\
Koodous & 4,175(2,390) & 573(351) & 139 & 34\\
Domain & 243(201) & 48(42) & 31 & 12\\
Security Reports & 84 (67) & 82 (65) & 28 & 22 \\
\hline
Total & 4,322(2,500) & 611(370) & 145 & 40\\
\noalign{\smallskip}\hline
\end{tabular}
}
\label{tab:malwaredataset} 
\end{table}

\section{General Overview}
\label{sec:trends}

\subsection{Dataset Overview}

We aggregate all the apks collected from the four channels and de-duplicate them according to the apk hash (SHA256). Finally, we collect a total of 4,322 coronavirus-themed apk samples (2,500 unique apps), released from January 2020 to November 2020. 
Regarding maliciousness testing, there are 611 samples (370 apps) flagged by at least one engine on VirusTotal and these samples will be regarded as the malware in this paper. Table~\ref{tab:malwaredataset} shows the overall distribution of our dataset, where each of the four channels is presented in detail to give a better picture of the COVID-19 themed apps.

There is an overlap of apps collected by these channels, especially between Koodous and the other three channels. Specifically, 29 out of 48 apks released in the app markets are also available in Koodous and more of the apps obtained from the domains (62\%) and security reports (60\%) exist in Koodous as well. 
Notably, there was not a single malware found in the app markets, reflecting the rigorous review of COVID-19 related apps in the app markets during the outbreak. Nevertheless, hundreds of COVID-19 themed malicious apps were found in related domains and other hidden sources collected in Koodous.
In our dataset, there are 48 malware correlated with 37 COVID-19 related domains. Interested in their interactions, we further investigate the relationships between apps and domains and find that there are two major types: (1) \textit{downloading relationship}, i.e., the malicious app can be downloaded from the corresponding domain and (2) \textit{communicating relationship}, i.e., the malicious app communicates with the domain. In addition, for the 37 COVID-19 themed domains, 26 of them (70\%) are flagged as malicious by VirusTotal. The remaining of them are websites that provide APIs for COVID-19 related information and statistics (e.g., \url{https://corona.lmao.ninja/}), which could be integrated by any apps.
Apps obtained from security reports are basically analyzed by vendors or analysts who then reveal their threats and suspicious behaviors to the public. Thus apps from this source are almost all malicious, which is consistent with our VirusTotal scan results in our dataset.
Besides, in addition to the available meta-information of all apks, we also download their binary files relying on the premium services provided by Koodous and VirusTotal.

For apps with multiple versions (apks), it is interesting to know whether all versions of an app are malicious or benign. Thus we carry out a further check and it turns out that the vast majority of the apps with multiple versions have the same maliciousness among all their apks. However, there are 6 exceptions whose early versions of the apks are non-malicious while later versions are malicious. For example, there is an app called ``Coronavirus 2019-nCoV", whose early four versions return no maliciousness from scan results in VirusTotal, while the later two versions are flagged by 19 and 27 engines respectively.

\begin{figure}
\centering
\includegraphics[width=0.95\textwidth]{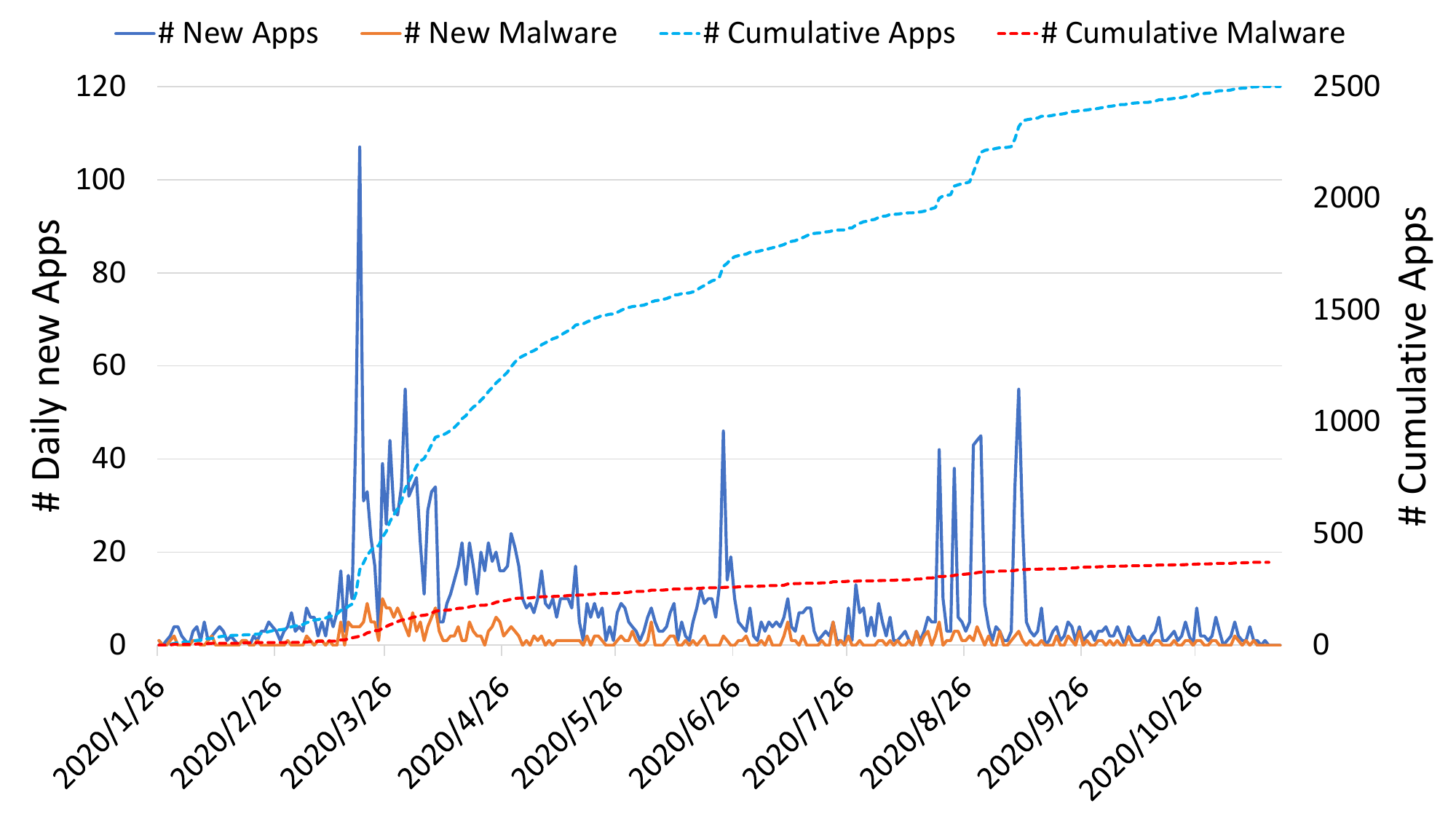}
\caption{Number of COVID-19 related apps and malware over the time (from Jan to Nov 2020).}
\label{fig:time-distribution} 
\end{figure}

\subsection{The trends of COVID-19 themed apps and malware}

For each app, we define its \textit{appear time} as the earliest time we found from various data sources.
For example, we have crawled the app upload time from Koodous, the app scan time (first and latest) from VirusTotal, and the app update time from app markets. The earliest one would be regarded as its \textit{appear time}. 
The distribution of the \textit{appear time} for the 2,500 COVID-19 themed apps\footnote{Note that, we did not consider different app versions here. For one app with multiple versions, we regard the time of the first version as its appear time.} and the 370 malware (with AV-rank $\geq 1$) is shown in Figure~\ref{fig:time-distribution}, where the number of daily new apps and the cumulative number are both presented.

The earliest app\footnote{app name: Avertisment Coronavirus H5N1,\\ MD5:bb3f343b219e7400551f04a1c17eb196} in our dataset was released on January 26, which is a COVID-19 themed ransomware upon our examination.
We can observe that the number of coronavirus related apps is quite low before March 15 (155 COVID-19 related apps and only 21 of them are considered to be malicious with AV-rank $\geq 1$). 
Subsequently, the number of COVID-19 related apps shows a rapid growth trend, where the largest number of COVID-19 related apps appeared on March 19, with 107 apps on that day. In addition, the highest number of COVID-19 related malware appeared on March 25 with 10 malicious apps. This coincides with the start of the global outbreak of COVID-19 and a sharp increase in the number of confirmed cases. Besides, the trend of increasing the number of COVID-19 related apps as well as malware becomes slower after April, 2020. This finding in part indicates the fact that the greatest interest and concern for COVID-19 is at the beginning of the outbreak.

\begin{framed}

\noindent \textbf{RQ \#$1$:} 
\textit{
There are over 4,300 COVID-19 related apks by the time of paper writing, and 611 of them are considered to be malicious (with AV-rank $\geq 1$). Most of them were released through channels beyond app markets, e.g., COVID-19 themed domains are used to distribute malware. Most of them were released after March 15, the time when the coronavirus became a pandemic. 
}

\end{framed}

\begin{figure}
\centering
\includegraphics[width=1\textwidth]{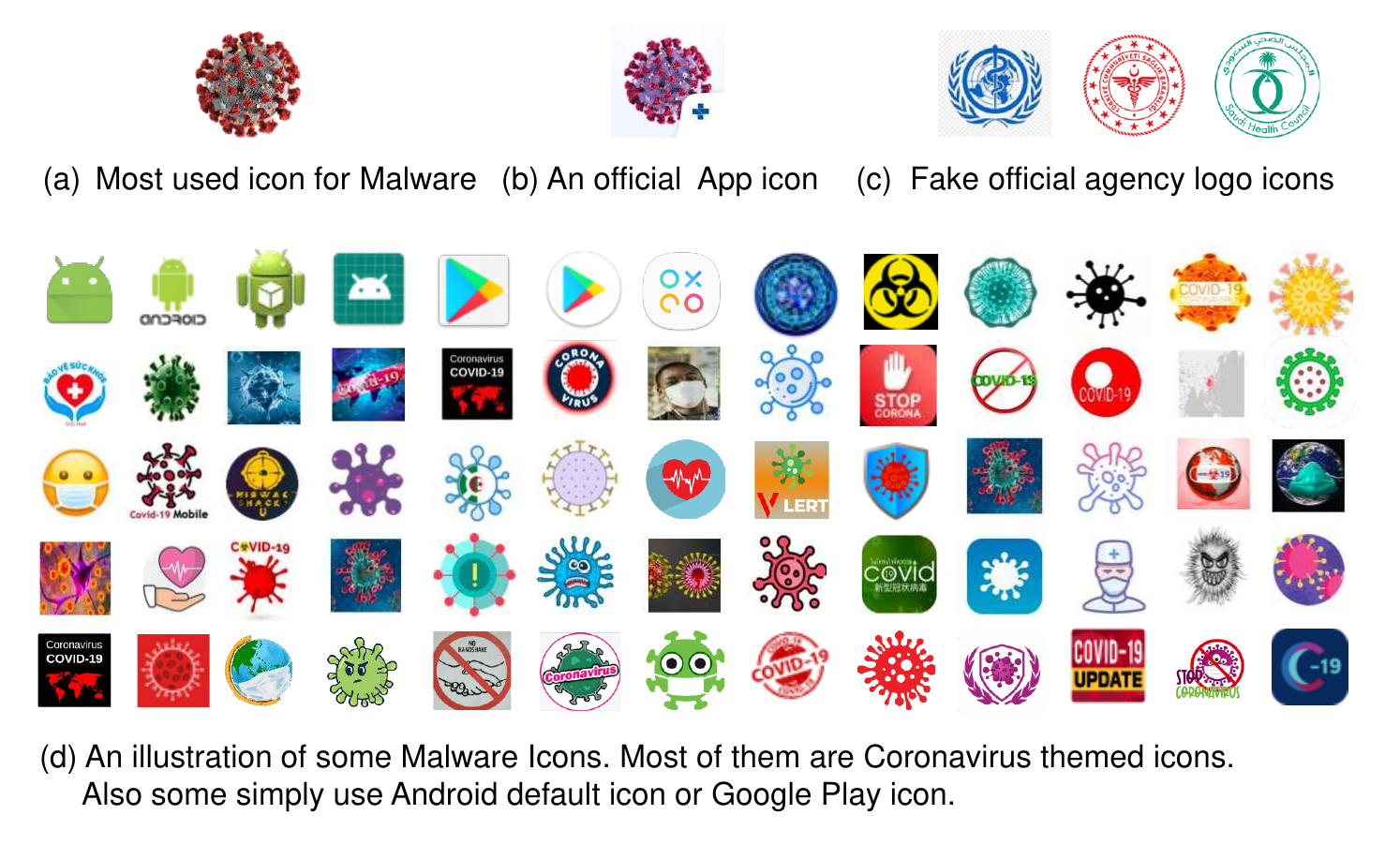}
\caption{The app icon of coronavirus-themed malware.}
\label{fig:app-icon} 
\end{figure}

\section{App Creation and Installation} 
\label{sec:creation}

We further investigate how these COVID-19 themed malicious apps are created and how they trick users to install them.
We consider two kinds of tricks here:
(1) \textit{fake apps}, and (2) \textit{repackaged apps},
which are the main social engineering based techniques\footnote{Social engineering is the term used for a broad range of malicious activities accomplished through human interactions~\cite{SocialEngineering-1}, which indicates an attempt by attackers to fool or manipulate humans into giving up access, credentials, banking details, or other sensitive information~\cite{SocialEngineering-2}. Fake apps and repackaged apps are believed to be common social engineering scams~\cite{App-SocialEngineering-1,App-SocialEngineering-2}.} to trick users into installing malicious apps based on previous studies~\cite{wang2018beyond,zhou2012dissecting,zhou2012detecting,hu2020mobile}.
A ``fake app'' masquerades as the legitimate one by mimicking the look or functionality. They usually have identical app names, package names or app icons to the original ones~\cite{kywe2014detecting,hu2020mobile}. 
While a “repackaged app” is one in which an attacker obtains a copy of the app from a distribution platform (e.g., Google Play) and thus typically shares most of the code with the original app. The attacker then adds malicious functionality (e.g., by decompiling the original app and inserting malicious payload), and re-distributes it to users who believe that they are using a legitimate app or the original app.

\begin{table}
\centering
\small
\caption{The main targets of fake apps.}
\resizebox{0.99\textwidth}{!}{
\begin{tabular}{|c|c|c|c|}
\hline\noalign{\smallskip}
App Name & Package Name & Downloads on GPlay & \# Fake apps\\
\noalign{\smallskip}\hline\noalign{\smallskip}

SM\_Covid19 & it.softmining.projects.covid19.savelifestyle & 100,000+& 16\\
\hline
COVID-19 & com.Eha.covid\_19 & 500,000+ & 15 \\
\hline
Stop COVID-19 KG & kg.cdt.stopcovid19 & 10,000+ & 10 \\
\hline
Coronavirus Help & appinventor.ai\_david\_taylor.Coronavirus\_help2020 & 5,000+ & 3 \\
\hline
Canada Covid-19 & ca.gc.hcsc.canada.covid19 & 100,000 + & 1\\
\noalign{\smallskip}\hline
\end{tabular}
}
\label{tab:fakeapp} 
\end{table}

\subsection{Fake Apps}
To quantify the presence of fake apps among our collection, our study was performed based on \textit{app identifiers} and \textit{app icons} respectively.

\subsubsection{Fake Apps with the Same App Identifiers.}
We take the following approach: if a malicious app shares the same app name or package name with an official/benign COVID-19 themed app in the dataset, but does not share a signature, we will regard it as a fake app. This approach is widely used in previous studies~\cite{wang2018beyond,wang2019understanding}.
In this manner, we have identified 303 fake app binaries out of 611 malicious apks (49.6\%). 
Most of the fake apps pretend to be the official ones for cheating. For example, 11 malware impersonate `SM\_Covid19'\footnote{package name: it.softmining.projects.covid19.savelifestyle}, an app launched by Italian government to assess the risk of virus transmission by monitoring the number, duration and type of contacts, by using the same app name and package name. 
Table~\ref{tab:fakeapp} lists five targeted apps available on Google Play that are mimicked by a number of fake apps.

\subsubsection{Fake Apps with Same/Similar App Icons.}
In some way, the app icon symbolizes a particular app as much as the app name or package name. Thus we further extract the icons of all the coronavirus-themed malware (with AV-rank $\geq 1$), and compare these icons with those of the official/benign apps to explore \textit{whether the attackers use icons to deceive users}. To be specific, if the icon of a malware is identical or extremely similar to the icon of an official/benign app, it is most likely that the malware is impersonating the corresponding normal one.
In this study, we take advantage of a near-duplicate image detector called DupDetector\footnote{https://www.keronsoft.com/dupdetector.html} and compare the malware icons to the set of normal app icons. This tool is proved to be effective in finding duplicate and similar images by comparing image pixel data, and is used in many other studies~\cite{KWON2014137,DAVIS2018224,https://doi.org/10.1002/jeab.176}. 

First, for the icons of all the normal apps, duplicates were eliminated using DupDetector and manual inspection. After that, we put all the malware icons into the system along with the benign app icons without duplicates for comparison, and selected a threshold of 98\% after several adjustments to ensure that there were no false positives for matching pairs. This method allowed us to detect 104 pairs of matching icons (a conservative estimate), from which we can determine that at least 104 malicious apps are trying to impersonate the corresponding normal apps by imitating the icons. The most used icon for malware is shown in Figure~\ref{fig:app-icon}(a), with over 30 malicious apps using this icon as well as some using its variants. We speculate that this icon is probably a parody of the original app Coronavírus - SUS\footnote{https://play.google.com/store/apps/details?id=br.gov.datasus.guardioes}, an official app released by the Brazilian government to notify the outbreak situation, whose icon bears a strong resemblance to it, as shown in Figure~\ref{fig:app-icon}(b).
In general, most malicious apps use coronavirus themed icons to induce users to download, which makes them appear more professional and credible. Also, some apps pose as other trusted organizations, e.g., some of them use the World Health Organization (WHO) logo as their icons to deceive users\footnote{An example app with MD5: 15e5a00c5d4ec8b4bbd0ebc70f0806aa}( Figure~\ref{fig:app-icon}(c)). Besides, there are a number of malicious apps that simply use Android's default icon and some use the Google Play icon.


\subsection{Repackaged Apps}
We further analyze how many of the malicious apps are repackaged from the official/benign apps, and whether the malicious developers reuse the same malicious payload to create a number of malware.

\subsubsection{Clustering based on App Similarity}
We first take advantage of the open source tool FSquaDRA2~\cite{EvaluationOfResourceBasedAppRepackagingDetection_Gadyatskaya2016} to measure the resource similarity of each app pair based on a feature set of resource names and asset signatures (the MD5 hash of each asset of an application excluding its icon and XML files), and then cluster them, which is widely adopted by previous studies~\cite{hu2019dating}.
FSquaDRA2 uses Jaccard distance to measure the similarity of two apps. Jaccard distance, also known as Jaccard similarity coefficient, is used to compare the similarity and difference between limited sample sets. The higher the Jaccard coefficient value, the higher the sample similarity. Based on the calculated results, we empirically set the similarity threshold as 90\% to cluster apps, i.e., for the apps with similarity scores higher than 90\%, we group them into the same cluster.
Note that, for apps with multiple versions (released by the same developer), we randomly leave one app during the app clustering phase.
In other words, each cluster contains at least 2 different apps (with different package name or developed by different developers).

\begin{figure}
\centering
\includegraphics[width=0.99\textwidth]{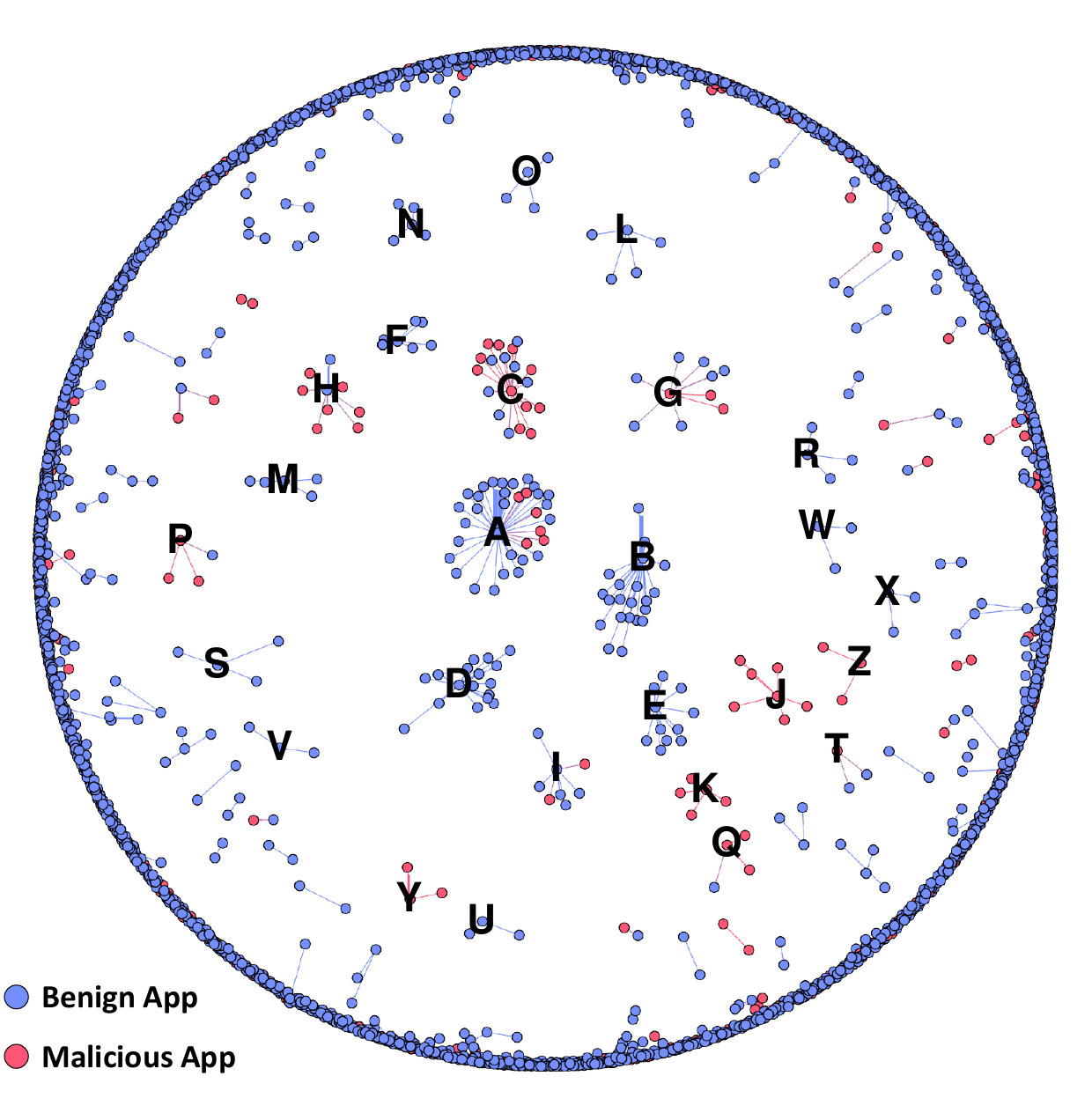}
\caption{App clustering result. Red node indicates malicious app and blue node indicates benign app. For each cluster, we randomly select one app and use edges to represent its similarity with other apps in the same cluster, i.e., the shorter the edge, the more similar they are. Note that isolated apps are shown in the peripheral circle.}
\label{fig:code_similarity} 
\end{figure}

Finally, we group 368 apps into 101 clusters and the remaining 1,945 are isolated apps (see Figure~\ref{fig:code_similarity}). 
Each node represents a coronavirus-themed app, where red node indicates the potential malware (with AV-rank $\geq 1$) and blue one indicates the benign app.
For each cluster, we randomly select one app and use edges to represent its similarity with other apps in the same cluster, i.e., the shorter the edge, the more similar they are.

\subsubsection{Result Analysis}

As a result, only 59 malicious apps have been grouped into 24 clusters. Note that, the clustered malware are not necessary to be repackaged malware from the benign/official COVID-19 apps. There are mainly two reasons. First, malware developers can release a number of apps with same/similar code and resource files, which can be identified by our clustering method (see Cluster K discussed in the following). Second, we observe that some apps with simple functionalities are created by taking advantage of popular app creation frameworks. Thus, they can be definitely grouped into clusters, while we should not regard them as repackaged malware (see Cluster A discussed in the following).

We further manually analyzed each cluster, and found that only a few malicious apps are actually repackaged. Among the 24 clusters containing malware, 9 of them contain only malicious apps, mainly because the malicious developers created a number of similar malware that share a lot of reused code and resources. Apparently, there is no repackaging phenomenon in these 9 clusters (containing 30 malware in total). For the remaining 15 clusters that include both benign and malicious apps, two of them (containing 9 malware in total) were clustered because the apps were created using the same app creation frameworks (AppsGeyser and App Inventor) while not because of repackaging.
Therefore, only 20 (5.4\%) malicious apps are identified as repackaged apps. 
In other words, \textit{most of the malicious apps are not repackaged based on existing COVID-19 benign apps}. This result is different with previous malware study~\cite{zhou2012dissecting} that over 80\% of malware samples are created based on app repackaging. 

To be more specific, we select three representative clusters (for each of the three scenarios respectively) for detailed description as follows. 

\textbf{Cluster A.} 
In cluster A, there are 34 coronavirus-themed apps and 6 of them are detected as malware. 
After our manual inspection, we fail to identify any repackaging in this cluster, but rather discover that the apps are quite similar because they are all created by the same app creation tool \texttt{AppsGeyser}\footnote{https://appsgeyser.com}, which is by far the biggest Free Android App Builder on the market, providing a step-by-step guide for creating Android apps. Their app names, package names and developer certificates are diverse while the format of package name is the same, i.e., prefixed with ``com.w" and followed by the app name. This finding indicates that there is not necessarily a repackaging relationship between malicious apps and benign apps even though they are clustered in a group, making the number of repackaged apps in our dataset actually smaller.


\textbf{Cluster K.} All the 5 apps in this cluster are detected as malware. 
We find that their package names are meaningless, which seems to be obfuscating, such as ``rnwjzlri.qiaopwnzcqrijy.ioyfsiukwf'', ``bqehgzgqygllillzks.lpugttk-ubu.erpwz dxnhtfmqwy'', etc. We further extract the developer certificates of these malware and find that these apps are signed by the same developer signature. However, the developer certificate\footnote{SHA1: 61ed377e85d386a8dfee6b864bd85b0bfaa5af81} is an Android common key and cannot be traced. As to their malicious behaviors, these malware samples use phishing window coverage and keystroke recording to steal victims' bank account information and credentials, which belong to the \textit{Cerberus} family, a well-known banking Trojan.

\textbf{Cluster P.} In this cluster, two out of three apps are detected as malware, which are indeed repackaged apps of the normal one after our scrutiny. Both malicious apps share the same app name, package name and icon as the benign app, but the developer signatures differ. We examine their requested permissions and discover that the original benign app requests only 3 permissions including ACCESS\_COARSE\_LOCATION, ACCESS\_FINE\_LOCATION and INTERNET. However, the repackaged app requests far more than that, with more than 20 permissions requested including 16 sensitive permissions such as READ\_CALL\_LOG, SEND\_SMS, WRITE\_EXTERNAL\_STORAGE, etc. We then drill down to the code level and perform manual dynamic analysis, and eventually identify its malicious payload which will download and install other malware onto the target endpoint device without the user's consent.

\begin{framed}

\noindent \textbf{RQ \#$2$:} 
\textit{
We investigate two main social-engineering based techniques (fake apps and repackaged apps) that are used by malware to trick users to install them. 
Many of the malicious apps (over 49.6\%) are camouflaged as official apps using the same app identifiers, and a number of them (over 100) use confusing similar app icons to mislead users. However, only a few of them (about 5.4\%) are repackaged from existing COVID-19 benign apps. 
}
\end{framed}

\section{Malicious Behaviors}
\label{sec:behavior}

As aforementioned, 611 COVID-19 themed apks are flagged by at least one anti-virus engine on VirusTotal (with AV-rank $\geq 1$). 
Table~\ref{tab:malware_avrank} shows top-10 of them ranked by the number of flagged engines. We next investigate the malicious behaviors of these 611 apks from \textit{malware category}, \textit{malware family} and \textit{anti-analysis techniques}.

\begin{table}[h]
\centering
\small
\caption{Top 10 COVID-19 themed malware with the highest malicious rank.}
\resizebox{0.99\textwidth}{!}{
\begin{tabular}{|c|c|c|c|c|}
\hline\noalign{\smallskip}
App Name & MD5 & AV-RANK & Family\\
\noalign{\smallskip}\hline\noalign{\smallskip}
Crona & 51902ba816f2b351d947419810a59f68 & 38 & spynote\\
Corona Safety Mask & d7d43c0bf6d4828f1545017f34b5b54c & 36 & piom\\
Corona1 & e5e97b95d4ca49d2f558169851af5eec & 36 & spynote\\
Covid19 & e8290dfcbe749bc8466bb886d805c49a & 34 & anubis\\
Covid19 & 6536f3ab0f70292e84d18413f86ca642 & 34 & anubis\\
corona mony & 7644e12134d86e558c59ffd1f063d447 & 33 & ransomkd\\
Covid19 & 9573615cd66921cb5f8c63b0e9bb764f & 33 & hqwar\\
Covid19 & 9abc81fda14ecc1abf8de278b852f521 & 33 & anubis\\
V-Alert COVID-19 & 439be2e754cfc5795d1254d8f1bc4241 & 33 & hqwar\\
COVID-19 & 199a0889756f3460cd634698803a280c & 33 & spynote\\
\noalign{\smallskip}\hline
\end{tabular}
}
\label{tab:malware_avrank} 
\end{table}

\subsection{Malware Category}
Malware can be classified into categories in order to distinguish the unique types of malware from each other. Classifying different types of malware is vital to better understand their malicious behaviors, the threat level they pose and how to protect against them~\cite{MalwareClassification}.
In this work, we follow the malware categories provided by Microsoft~\cite{microsoftcategory} for COVID-19 themed malware classification. 

\subsubsection{Malware Classification Method}
\begin{figure}
\centering
\includegraphics[width=0.95\textwidth]{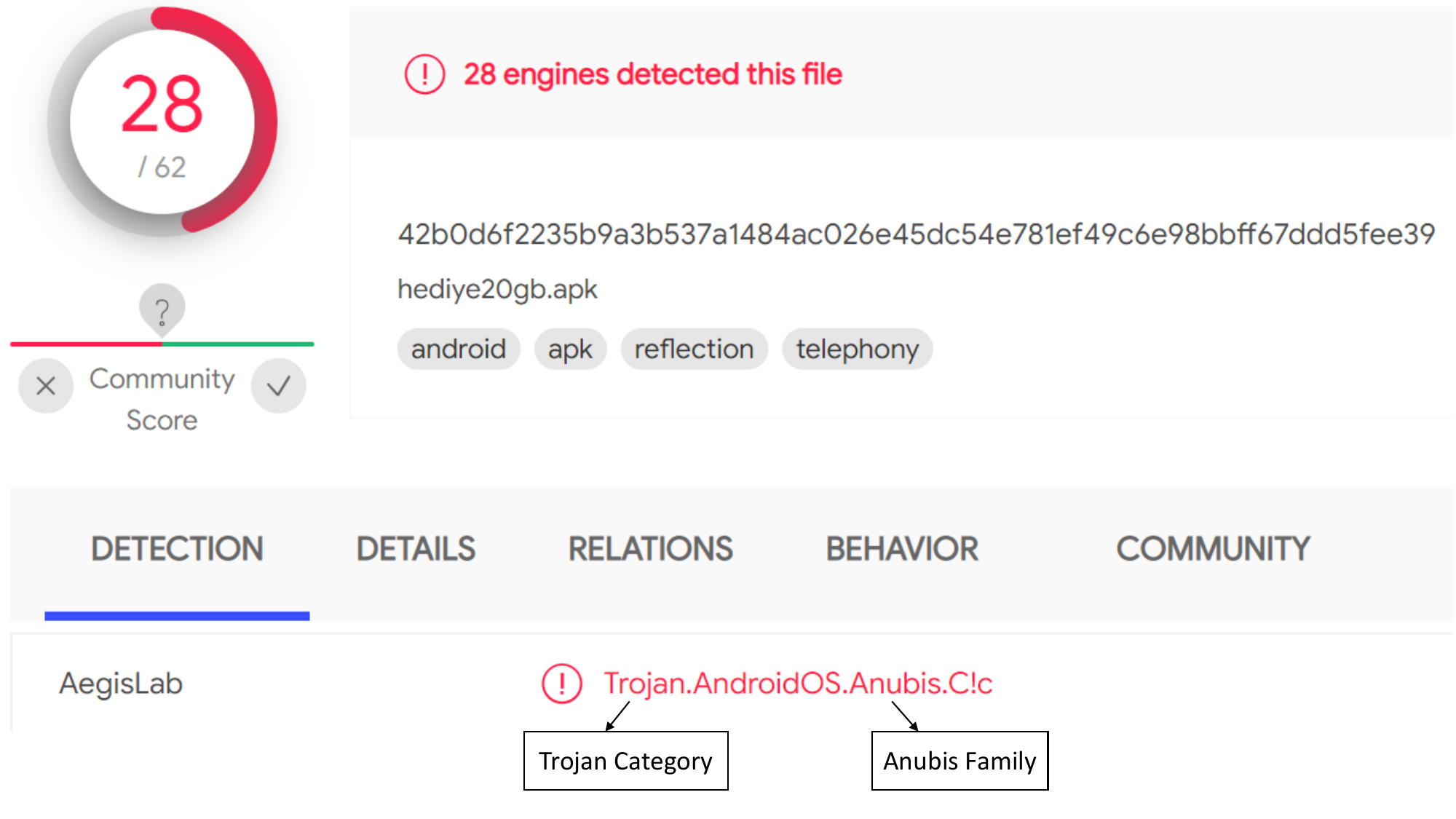}
\caption{An example of a malware label given by the VirusTotal engine.}
\label{fig:virustotal} 
\end{figure}

Based on the anti-virus labels (AV-labels) provided by VirusTotal and the family labels generated by AVClass (see Section 5.2), we have classified the malware into five main categories, including \textit{Trojan}, \textit{Ransomware}, \textit{Adware}, \textit{Riskware} and \textit{Spyware}. Our method has the following two steps: First, search the family name obtained from AVClass via Google, some of which can be searched directly to the category in which the family is located. For example, when searching `hqwar' on Google, we can clearly understand that it is a Trojan.
Second, according to the labels of each sample given by VirusTotal, we count the number of labels containing each category wordings, and choose the one with the largest number as its category. Figure~\ref{fig:virustotal} shows an example of a VirusTotal scanning label for a malware, through which we can observe that the sample belongs to `Trojan' category and `Anubis' family. Although the presentation format of the scan results varies slightly from engine to engine, we can identify which category is being reported by keyword matching. For example, there is an apk sample called `COVID-19'\footnote{MD5:6197187acd057f6bb4be25808fe3c8a8}, and there are 19 engines in VirusTotal giving malicious tags. The category keyword that is mostly contained in these tags is `Trojan' with 11 tags, thus we classify it into \textit{Trojan} category. By this method, we classify the malware in our dataset into these five categories.

\subsubsection{Result}

We next present the percentages and details of each category.

\begin{figure}
\centering
\includegraphics[width=0.99\textwidth]{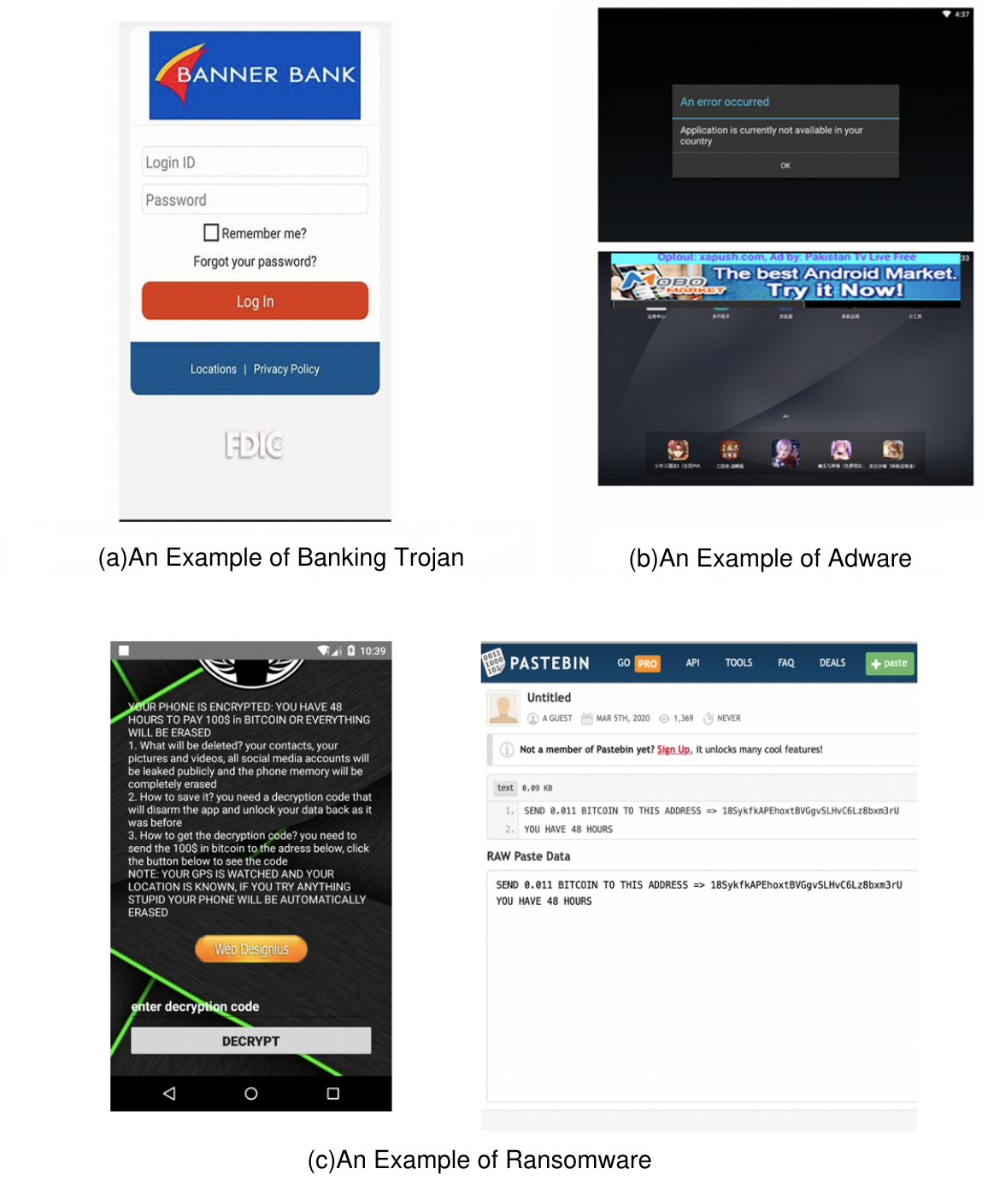}
\caption{Examples of Covid-19 themed malware.}
\label{fig:app_sample} 
\end{figure}

\textbf{Trojan (56\%)} Trojans that run on the Android operating system are usually either specially-crafted programs that are designed to look like desirable software, or copies of legitimate programs that have been repackaged or trojanized to include harmful components. 
For example, the malware\footnote{MD5: b7070a1fa932fe1cc8198e89e3a799f3} (app name: CORONA TAKIP) is a banking Trojan targeting Turkish users that belongs to the `Anubis' family. This malware disguises as an app to provide coronavirus information. 
However, it requires excessive permissions when it is installed and activated. Furthermore, it shows a phishing user interface (i.e., a bank login UI) at runtime to steal the victim's bank account, as shown in Figure~\ref{fig:app_sample} (a).

\textbf{Riskware (4\%)} Riskware is created by malicious developers to delete, block, modify or copy the victim's data, and destroy the performance of the devices or the network. For example, the malware\footnote{MD5: eca383edee4ef0db4961fc26db3d35b4} (app name: Covid-19 Visualizer) is detected as `Fakeapp' family, which disguises as a normal app that provides real-time query of the COVID-19 outbreak. Once launched, the malware will remind users to install the ``Adobe Flash'' plugin to display the full content. After obtaining user authorization, the malware will run in the background, leaking users' privacy and intercepting phone calls and SMS messages.

\textbf{Spyware (29\%)} Spyware is apps that record information about mobile users or what mobile users do on their phones without users' knowledge. RAT (remote administration tool) is a kind of popular spyware on Android, and there are a number of RAT frameworks that can be used to create spyware.
Android Spyware usually collects victims' private data, call records, message records and photos and sends them to the hackers secretly. For example, the malware (package name: kg.cdt.-stopcovid19\footnote{MD5: 6588e22e9c9d35179c166113d3de325b}) is detected as `Datacollector' family, which steals the users' personal privacy and sends it to the attacker.

\textbf{Adware (4\%)} Adware is a form of malware that hides on a user's device and serves aggressive (or fraudulent) advertisements. Some adware also monitors users' behavior online so it can target users with specific ads. For example, the adware named \emph{``Coronavirus Tracker''\footnote{MD5: e423f61f1414eccd38649f20d018723d}} is detected as the `Hiddenads' family. Once launched, it informs the user ``not available in your country'' and uninstall itself. Actually, it just hides the app icon and keeps running in the background. The malware pops up some aggressive advertisements at intervals, as shown in Figure~\ref{fig:app_sample} (b).

\textbf{Ransomware (7\%)} Once launched, the ransomware will lock the victims' devices or files and force the users to pay a ransom to protect their important data. As shown in Figure~\ref{fig:app_sample} (c), the malware\footnote{MD5: 69a6b43b5f63030938c578eec05993eb} disguises as the Coronavirus Tracker app to provide information about the COVID-19. In fact, it is a ransomware that locks the victims' files and asks for Bitcoin. Specifically, the Bitcoin address\footnote{BTC address: 18SykfkAPEhoxtBVGgvSLHvC6Lz8bxm3rU} is not hard-coded in the APK file. Once clicking the button shown on the locking UI, it will redirect users to an external webpage that shows the real Bitcoin address.

\begin{table}
\centering
\small
\caption{A Summary of Malicious behaviors of the COVID-19 themed malware (classified by malware family): Privacy Stealing (Privacy), SMS/Phone Calls (Phone), Remote Control (Control), Bank Stealing (Bank), Ransomware (Ransom), and Aggressive Advertising (Ads).}
\resizebox{0.99\textwidth}{!}{
\begin{tabular}{|c|c|c|cccccc|}
\hline\noalign{\smallskip}
Family Name& \# Malware & Malware Category&\multicolumn{6}{|c|}{Malicious Behaviour}\\
&&&Privacy&Phone&Control&Bank&Ransom&Ads\\
\noalign{\smallskip}\hline\noalign{\smallskip}
hqwar& 44 & Trojan &$\surd$& &$\surd$&$\surd$& & \\
spynote & 39 & Spyware &$\surd$& & & & & \\
anubis & 38 & Trojan &$\surd$&$\surd$&$\surd$&$\surd$& & \\
timethief & 19 & spyware &$\surd$& &$\surd$& & & \\
cerberus & 17 & Trojan &$\surd$& &$\surd$&$\surd$& &\\
masplot & 13 & Trojan &$\surd$& &$\surd$& & &$\surd$ \\
metasploit & 7 & Ransom &$\surd$&&$\surd$&&$\surd$&\\
dnotua& 5 & Trojan &$\surd$&$\surd$&$\surd$&&&\\
locker & 5 & Ransom &$\surd$&&$\surd$&&$\surd$&\\
boogr & 5 & Trojan &$\surd$&&&&&\\
spymax & 4 & Trojan &$\surd$& &$\surd$& &$\surd$& \\
utilcode & 4 & Spyware &$\surd$&&&&&\\
xploitspy & 4 & Spyware &$\surd$&&$\surd$&&&\\
hiddad & 4 & Adware &$\surd$&$\surd$&$\surd$&&&$\surd$\\
fakeapp & 4 & Riskware &$\surd$&$\surd$&$\surd$&&&\\
covidspy & 3 & Spyware &$\surd$&$\surd$&$\surd$& & & \\
ransomkd & 3 & Ransom &$\surd$&&$\surd$&&$\surd$&\\
bodegun & 3 & Adware &$\surd$&&&&&$\surd$\\
pigetrl & 3& Trojan &$\surd$& & & & & \\
ahmyth & 2 & Spyware&$\surd$&$\surd$&$\surd$&&&\\
notifyer & 2 & Adware &$\surd$&$\surd$& & & & \\
piom & 2 & Trojan &$\surd$&$\surd$&$\surd$&&&\\
lockscreen & 2 & Ransom &&&$\surd$&&$\surd$&\\
jiagu & 2 & Riskware & & & & & &$\surd$ \\
hiddenapp & 2 & Rishware &$\surd$&&$\surd$&&&\\
ewind & 1 & Adware & &$\surd$&&&&$\surd$\\
hiddenads & 1 & Adware & & & & & &$\surd$\\
mobidash & 1 & Riskware & & & & & &$\surd$\\
apkprotector & 1 & Trojan &$\surd$& &$\surd$& & & \\
svpeng & 1 & Trojan &$\surd$&$\surd$&$\surd$&$\surd$&$\surd$&\\
fakeinst & 1 & Trojan &$\surd$&$\surd$& & & & \\
trakcer & 1 & Trojan &$\surd$&$\surd$&$\surd$&$\surd$& & \\ 
dingwe & 1 & Trojan &$\surd$& &$\surd$& & & \\
bankbot & 1 & Trojan &$\surd$& &$\surd$& &$\surd$& \\
uten & 1 & Trojan &$\surd$&&&&&\\
sandr & 1 & Spyware &$\surd$&&&&&\\
smsthief & 1 & Trojan &$\surd$&$\surd$&$\surd$& & &$\surd$ \\
homeproxy & 1 & Trojan &$\surd$& &$\surd$& & & \\
resharer & 1 & Riskware &$\surd$&$\surd$&$\surd$& & & \\
traca & 1 & Spyware &$\surd$& &$\surd$& & & \\

\noalign{\smallskip}\hline
\end{tabular}
}
\label{tab:malware_family} 
\end{table}

\subsection{Malware Family}
\label{section:malware fmaily}

\subsubsection{Family Labelling Method}
\begin{figure}
\centering
\includegraphics[width=0.95\textwidth]{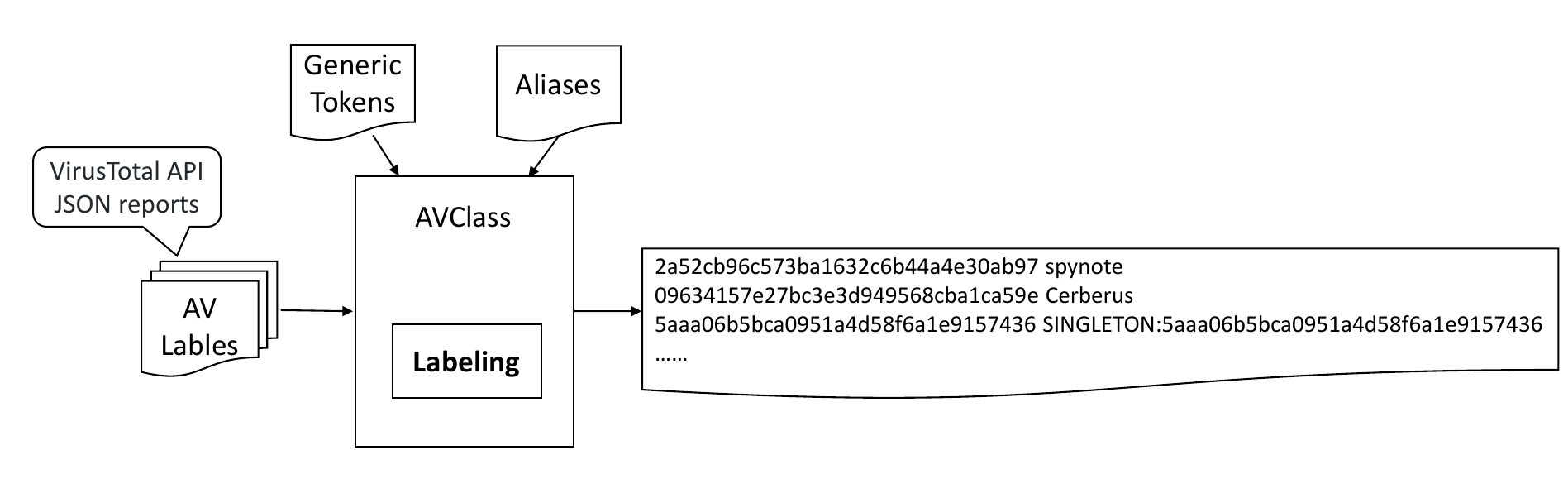}
\caption{Label the malware family name through AVClass.}
\label{fig:workflow} 
\end{figure}

A malware family is a group of applications with similar attack techniques~\cite{MalwareFamily}. Identifying the malicious families is a complex process involving the categorization of potentially malicious code into sets that share similar features, while being distinguishable from unrelated threats or non-malicious code~\cite{6107902}.
We leverage AVClass~\cite{sebastian2016avclass}, a widely used malware family tagging tool to label the malware family name for each sample. Give it as input the AV labels for a large number of malware samples (e.g., VirusTotal JSON reports) and it outputs the most likely family name for each sample that it can extract from the AV labels. It can also output a ranking of all alternative names it found for each sample. In a nutshell, AVClass comprises two phases: preparation (optional) and labeling. For our work, we are only interested in the labeling, which outputs the family name for the samples. Figure~\ref{fig:workflow} illustrates our workflow. We take as input VirusTotal API JSON reports with the AV labels of malware samples to be labeled, a list of generic tokens and a list of aliases (default). It outputs the most likely family name for each sample as shown in Figure~\ref{fig:workflow}, which means sample `2a52cb96c573ba1632c6b44a4e30ab97' is most likely from the `spynote' family and `09634157e27bc3e3d949568cba1ca59e' from the `Cerberus' family. The third line means that for sample `5aaa06b5bca0951a4d58f6a1e9157436' no family name was found in the AV labels. Thus, the sample is placed by himself in a singleton cluster with the name of the cluster being the sample's hash. 

\subsubsection{Result}
In this way, these malicious apps are classified into 40 families in total. Note that, AVClass is unable to label all the flagged apps with family names and there are 250 apks labelled in our study.
As shown in Table~\ref{tab:malware_family}, we list the distribution of all the malware families. 

For each malware family, we randomly select two apps from our dataset (if there are more than two apps in this family) and perform manual examination to label their malicious behaviors. 
Our manual analysis consists of two parts:
(1) \textit{Static analysis}. Our static analysis includes extracting the declared permissions and component information from the \texttt{Manifest} file, analyzing the embedded third-party libraries based on LibRadar~\cite{ma2016libradar}, pinpointing the sensitive API invocation, and analyzing the sensitive information flow using FlowDroid~\cite{arzt2014flowdroid}. 
Based on these information, we can know whether the malicious apps perform SMS/CALL related activities, invoke aggressive advertising libraries, release private information, and other sensitive behaviors.
(2) \textit{Dynamic analysis}. We first install these apps on the real smartphone, and check their behaviors by interacting with them using both DroidBot~\cite{li2017droidbot} (a widely used automated testing tool for Android) and manually clicking. During runtime, we can check whether the malicious apps show aggressive and annoying advertisements, redirect users to malicious and fraudulent websites, lock users' phones, etc. 
Besides, we have recorded all the network traffic to check whether the malware communicates with the remote servers.

\begin{figure}
\centering
\includegraphics[width=0.95\textwidth]{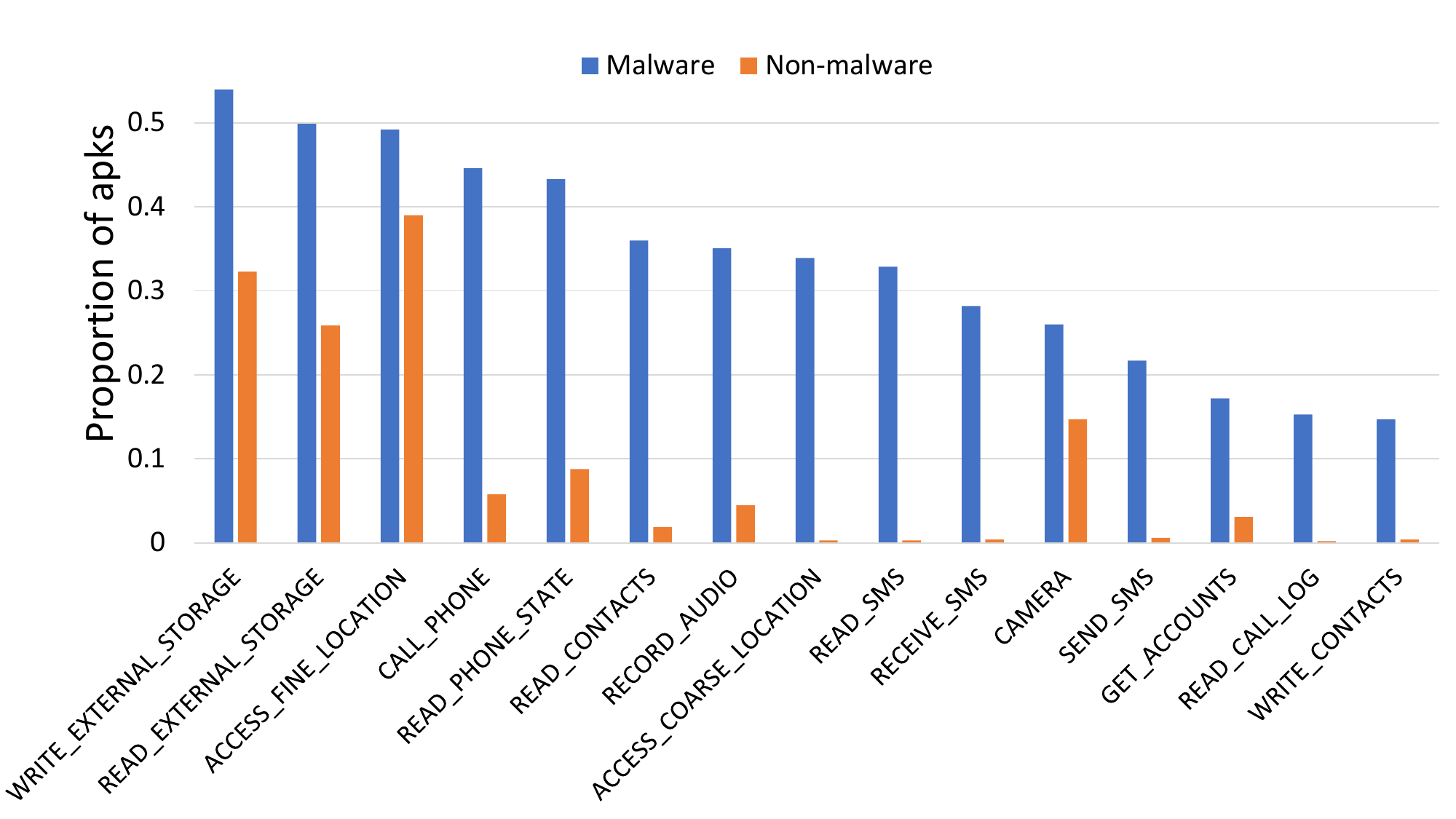}
\caption{Sensitive permission requests for COVID-19 themed malware and non-malware.}
\label{fig:permission} 
\end{figure}

Based on the aforementioned exploration, we have classified the malicious behaviors into six major categories (see Table~\ref{tab:malware_family}), including \textit{Privacy Stealing}, \textit{Sending SMS/Phone Calls}, \textit{Remote Control}, \textit{Bank Stealing}, \textit{Ransomware}, and \textit{Aggressive Advertising}.

\begin{itemize}
    \item Most of the COVID-19 related malware families have \textbf{privacy stealing behaviors}, i.e., 35 out of 40 families steal users' personal data without declaring the proper purposes of permission use. To be specific, we investigate and compare the behaviors of COVID-19 malicious apps and benign apps requesting sensitive permissions. Figure~\ref{fig:permission} shows the top-15 sensitive permissions used in the malware versus their usage in non-malware, including ``Write/Read External Storage",``Call Phone", ``Read Phone State", ``Read/Write Contacts", ``Access FINE/COARSE Location", ``Read/Receive/Send SMS", etc. It is surprising to see that, the percentage of malicious apps requesting these privacy permissions is much higher than non-malware on almost every permission, revealing a huge risk of malware stealing users' private information. Besides,
    some malicious apps even use the sensitive permissions that are only available in the latest Android SDK versions. For example,  ACCESS\_BACKGROUND\_LOCATION and ACTIVITY\_RECOGNITION are introduced in API level 29 (Android 9.0), which allow an app to access location in the background and recognize physical activities, respectively.
    \item \textbf{Remote control} is the second largest behavior category. These malicious apps communicate with remote C\&C servers and receive commands from the servers to perform related malicious actions and send the collected data to the attacker. We have identified 27 families that receive commands from remote servers.
    \item Roughly 35\% of malware families have the behavior of \textbf{sending text messages or making phone calls}. These malware send high-rate SMS messages, make phone calls or subscribe without user authorization to obtain financial benefits.
    \item Besides, five families steal users' banking information. The malicious developers carefully design a phishing page similar to the official bank login or payment interface to confuse the victim, or redirect to a third-party website when the user performs a bank operation. 
    \item There are seven families that are indeed ransomware that ask for Bitcoins to make a profit. Once launched, it will encrypt the victim’s mobile phone files or force a lock screen and extort a high ransom.
    \item Furthermore, we have identified eight aggressive adware families exploited by COVID-19 themed malware. Once launched, the adware will pop up full-screen ads every once in a while and auto-click to display the ads, which seriously affects the normal use of users.
\end{itemize}

\subsection{Anti-analysis Techniques}

Previous work~\cite{wang2017droid,rasthofer2016harvesting} suggested that sophisticated malicious apps have exploted a number of anti-analysis techniques to evade detection.
Thus, we further seek to analyze whether and to what extent the COVID-19 themed malicious apps have such behaviors. 

\subsubsection{Method}

In this work, we take advantage of APKid~\cite{apkid}, an Android App Identifier for packers, protectors, obfuscators and oddities, to detect whether the packers, obfuscators, and other anti-analysis techniques are used in COVID-19 themed apps. APKiD can look at an Android APK or DEX file and detect the fingerprints of several different compilers.
We feed all the 611 malicious apk samples to the APKid Identifier. It will decompose APKs and try to find compressed APKs, DEX, and ELF files, and output the anti-analysis techniques used in the apk. For comparison, we analyze the anti-analysis techniques used in the remaining 3,711 non-malicious apps as well. 

\subsubsection{Result}

\begin{figure}
\centering
\includegraphics[width=0.9\textwidth]{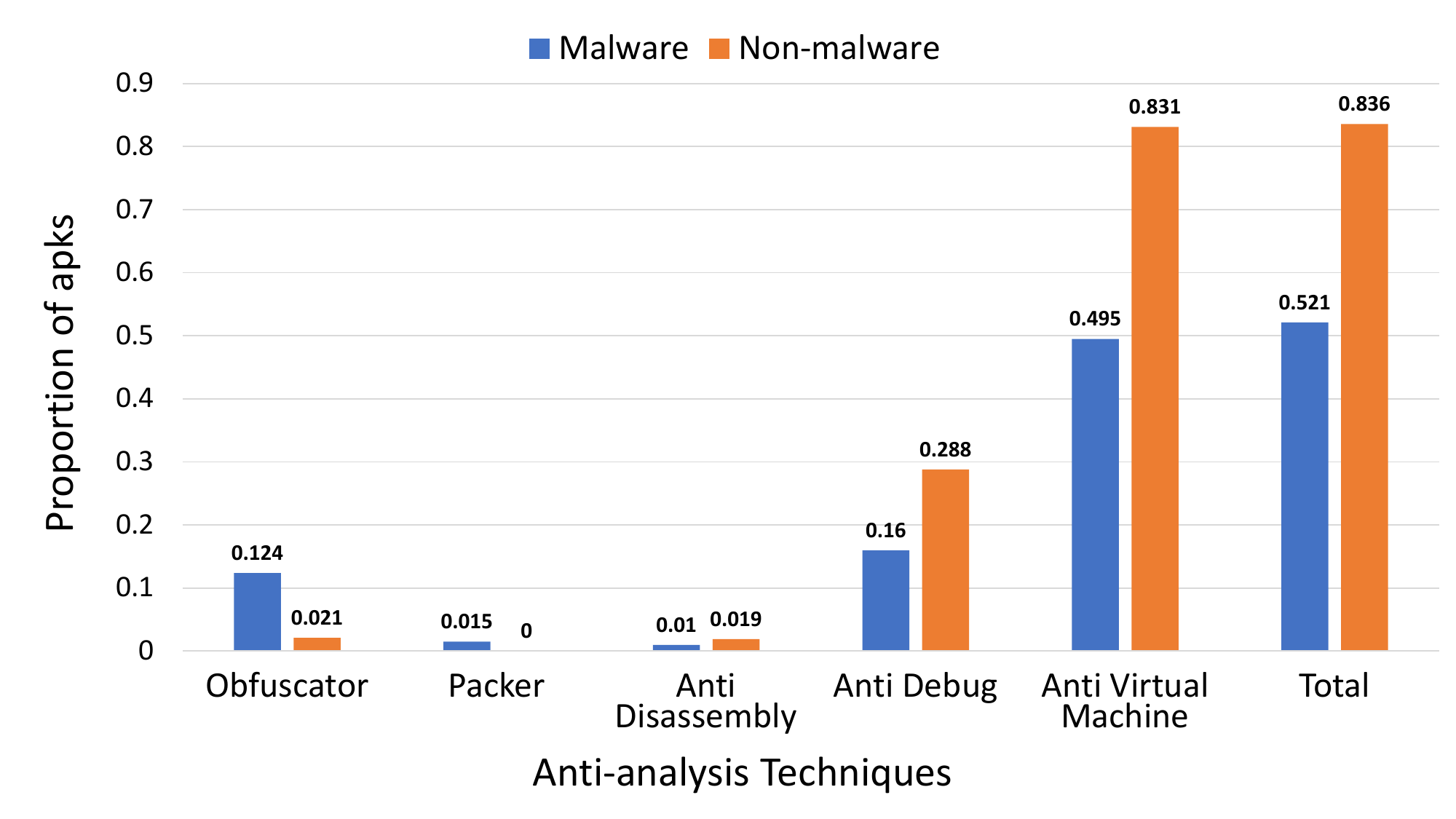}
\caption{Usage of anti-analysis techniques for COVID-19 themed malware and non-malware.}
\label{fig:anti-analysis} 
\end{figure}

Based on the experiment results, we classify the anti-analysis techniques used by COVID-19 themed malware into the following five categories and the proportion of each technique being used is shown in Figure~\ref{fig:anti-analysis}. Overall, 52.1\% of COVID-19 themed malware have adopted anti-analysis techniques, while surprisingly, the percentage is 83.6\% in COVID-19 themed benign apps. 
This result suggests that such malicious apps do not show stronger motivation than benign apps to protect their apps from being analyzed. The possible reason is that many of the benign COVID-19 themed apps are created by the official organizations (e.g., the government). Thus, they take actions to protect themselves from being hacked by malicious developers. 
Moreover, a large number of the benign COVID-19 themed apps are related to contact tracing, exposure notification, etc., which can only be operated on real smartphones. Thus, over 80\% of them will check whether they are running on the smartphones or virtual machines (using anti virtual machine technique). Nevertheless, as to the code obfuscation and packer, malicious apps that adopted anti-analysis techniques are much higher than that in benign apps, indicating that there is indeed some malware that exploits these anti-analysis techniques to try to evade detection.
Besides, we compared some benign apps and repackaged version of the same apps for anti-analysis and found that the repackaged apps using the same anti-analysis as the original ones, without adding new means. This indicates that the repackaged apps basically follows the original anti-analysis techniques and do not take much effort in this area.
We next present the detailed descriptions for each type of anti-analysis techniques.

\begin{itemize}
    \item \textbf{Code Obfuscation.} Obfuscation is the process of modifying an executable APK file. It modifies actual method instructions or metadata, and it does not alter the output of the program. Obfuscator includes renaming strings, variables and method names, encrypting data, etc. It makes the decompiled source code more difficult to understand, and makes it more difficult for security personnel to analyze malicious apps. In our dataset, there are 12.4\% COVID-19 themed malware that use obfuscation techniques, a much higher percentage compared to the 2.1\% for the benign apps.  
    \item \textbf{Packer.} In order to strengthen the protection of Android, the app pack Dex files to prevent them from being cracked by static decompilation tools and leaking the source code. For example, the malware\footnote{MD5:3c0b5bc0ef6b143e51be7f3cd0028994} (app name: corona viruse) use ApkProtect~\cite{ApkProtect} tool to pack the Apk file. There are only 9 malware that employ this technique and none of normal apps use it in our dataset. 
    \item \textbf{Anti Disassembly.} The apk file is actually a zip package. We can disassemble the apk files, and decompile them to obtain the resource files and source code. Anti-disassembly technique is to prevent the apk file from being disassembled. Anti-disassembly uses specially crafted code or data in a program to cause disassembly analysis tools to produce an incorrect program listing. For example, the malware\footnote{MD5: e521b0e519c0f08217e3e90c894f8094} (app name: Corona Updates) adds code segments with illegal class names, which invalidates the decompilation tools. In our dataset, the number of apps that use anti-disassembled technique is quite low for both malware and non-malware.
    \item \textbf{Anti Debug.} Malicious apps can avoid some dynamic debugging techniques by listening to port 23946 (default port of android\_server) and debugging related processes such as android\_server, gdb, gdbserver, etc. In our dataset, there are 16\% malware samples use the \texttt{Debug.isDebuggerConnected()} method to check whether they are in debugging. While for the normal apps, the percentage using anti-debug is a bit higher, about 29\%.
    \item \textbf{Anti Virtual Machine.} The malware check whether they are running on real devices by analyzing the environment in which the APK runs, checking device information, device serial numbers, sandbox processes, feature directories and files of the simulator, etc. Once it is detected that it is not running on a real device, some malicious behavior will not be triggered to avoid dynamic detection. About half of the COVID-19 themed malware detect the running environment, sandbox processes, and device hardware serial numbers to avoid analysis. While, this technique is more extensively used in COVID-19 related non-malware, with roughly 83\% of normal apps using anti virtual machine (anti-VM) technique.
\end{itemize}

\begin{framed}

\noindent \textbf{RQ \#$3$:} 
\textit{
Trojan and Spyware are the two main categories for COVID-19 themed malware. Their purposes are either stealing users' private information, or making profit by cheating users using tricks like phishing pages, sending premium SMS/Phone calls, stealing bank accounts, and locking the phones. Anti-analysis techniques have been used by roughly 52\% of COVID-19 related malware, compared to 83\% of the non-malware.
}
\end{framed}

\section{Malicious Actors}
\label{sec:developers}

Our aforementioned study indicates the prevalence and characteristics of COVID-19 related Android malware. We next seek to understand the malicious actors behind them. 

\subsection{The Developers of COVID-19 Malware}

Generally, a unique developer certificate can be used to identify the developer.
We found that some malware developers may use the known common keys in the community to sign apps.
The most famous keys are the publicly known
private keys included in the AOSP project. The standard
Android build uses four known keys, all of which can be found at
\texttt{build/target/product/security}. For example, \textit{TestKey} is the generic default key for packages that do not otherwise specify a key. Other publicly-known keys include \textit{Platform (key)}, \textit{Shared (key)} and \textit{Media (key)}. Thus we collect these keys and compare them with the signatures we extracted, and two of them were identified. After our inspection, over half of the malicious apps signed by the generic keys. To be specific, 51\% of them were signed by \textit{TestKey}. It suggests that malicious developers tend to hide themselves, without using private signatures. However, previous study~\cite{wang2019characterizing} revealed security issues for apps signed with the publicly-known keys that it is easy for attackers to replace the vulnerable app with another one (possible with malicious payloads), without user’s knowledge. Thus developers of benign apps generally use their own private keys to prevent such security problems.
It also implies that app markets should regulate the usage of generic signatures, and mobile users should pay special attention to apps signed by these common signatures (during the app installation phase, the Android system usually prompts the developers of the apps).

For other developer signatures, we further search them on Google to confirm they are not publicly known signatures. 
At last, we have 145 private signatures left. 34 of them have created more than one malicious app in our dataset, with a maximum of 29 malware from one developer signature.

\begin{table}
\centering
\small
\caption{The top 10 habitual malicious developers.}
\resizebox{0.99\textwidth}{!}{
\begin{tabular}{|c|c|c|c|c|cc|}
\hline\noalign{\smallskip}
\multirow{2}*{Developers Certificate (SHA1)}&\multirow{2}*{Earliest Active Time} & \multirow{2}*{Country Code}&\multirow{2}*{\# COVID-19 Malware}  & \multirow{2}*{\# Released Apps}&\multicolumn{2}{c|}{\# Malware}\\
&&&&&AV-Rank $\geq$ 1&AV-Rank $\geq$ 10\\
\noalign{\smallskip}\hline\noalign{\smallskip}
ece521e38c5e9cbea53503eaef1a6ddd204583fa & 2014-10 & ID (Indonesia) & 3 & 83148 & 99\% & 93\%\\
b79df4a82e90b57ea76525ab7037ab238a42f5d3 & 2014-10 & US & 3 & 34537 & 42\% & 15\%\\
927ca44949d7788aa86f9d7f04d7fdacecd1dfb9 & 2016-02 & None & 16 & 16668 & 27\% & 6\%\\
496d0c5ab97813582969b42d497124b26583ddf7 & 2014-10 & CN (China) & 1 & 15973 & 69\% &13\%\\
6d2aa36c370d8b6156dba70798a8c6c728265404 & 2015-10 & IN (India) & 3 & 13718 & 88\% &67\%\\
55a48e1a17a067c7fb22efb3639558eac0fc313f & 2014-10 & None & 11 & 12701 & 48\% & 10\%\\
09dceb70d91de79335b6c143d05f9a6b6de9e59c & 2019-08 & IT (Italy) & 2 & 8806 & 100\% & 97\%\\
b0ce633eae17195c31325c74e33e3bb90482076d & 2014-10 & US & 1 & 6555 &78\% &6\%\\
219d542f901d8db85c729b0f7ae32410096077cb & 2014-11 & AU (Australia) & 4 & 6397 & 95\% &52\%\\
dd2b8fab67577ce55712d9881deba7e76d7b8df5 & 2016-10 & US & 6 & 5792 & 86\% & 16\%\\
\noalign{\smallskip}\hline
\end{tabular}
}
\label{tab:developers}
\end{table}

\textbf{Habitual Malicious Developers.}
We hypothesize that, these malicious apps are created by habitual malicious developers, and they just take advantage of the coronavirus pandemic to lure unsuspicious users. To verify our hypothesis, we seek to collect more apps released by these developers.
Thus, we take advantage of Koodous to crawl all the apps released by these 145 malicious developers. Finally, we harvest 228,443 apps in total.
We further check all the detection results of these apps from VirusTotal. 

\begin{figure}[t]
\centering
\includegraphics[width=0.95\textwidth]{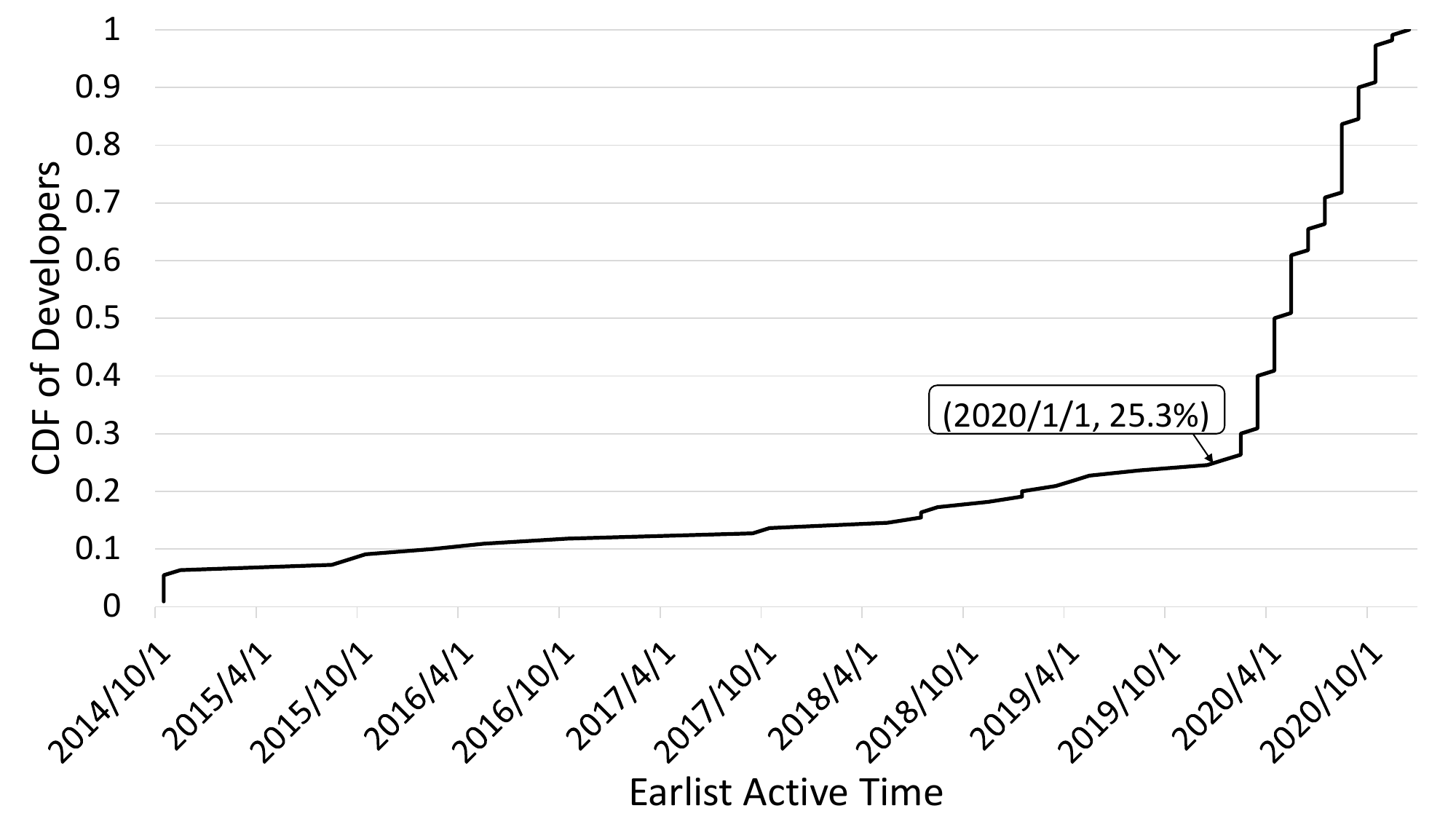}
\caption{CDF of the earliest active time of the developers.}
\label{fig:pdeveloper} 
\end{figure}

As shown in Figure~\ref{fig:pdeveloper}, about a quarter of malicious developers have released at least one app before the COVID-19 outbreak. Table~\ref{tab:developers} shows the top-10 habitual developers ranked by the number of released apps. Some of them are popular since 2014.
However, from another point of view, \textit{roughly 75\% of the COVID-19 malware developers are emerging developers that target this pandemic}, 
This observation contradicts our hypothesis. 

\begin{figure}
\centering
\includegraphics[width=0.95\textwidth]{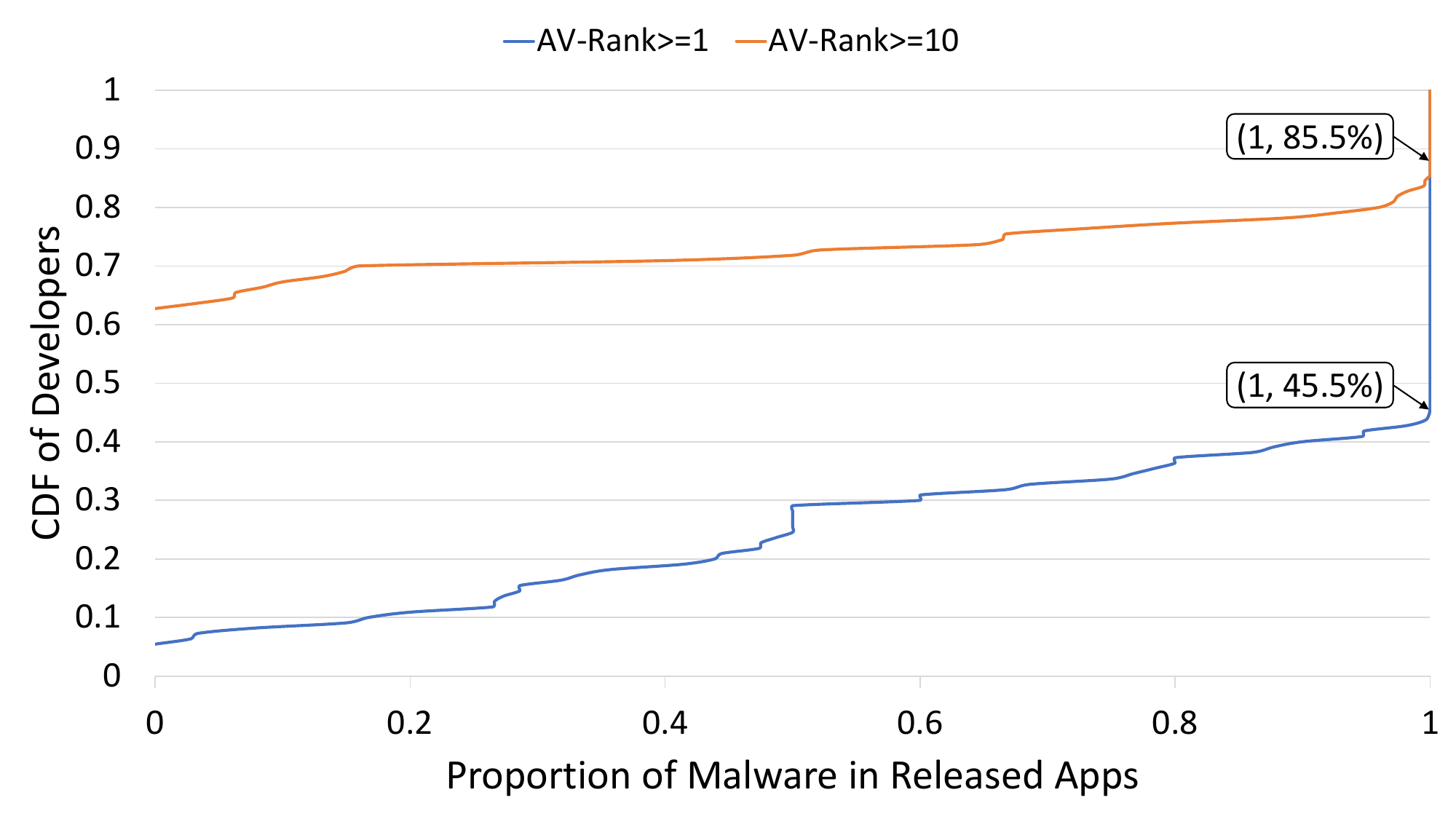}
\caption{CDF of proportion of malware released by developers.}
\label{fig:maldeveloper} 
\end{figure}

We further investigate whether these developers are focused only on creating malware, by calculating the proportion of malware samples among all the apps they developed (defined as \textit{Malware Rate}). Here, we have adopted two thresholds to flag a malware, i.e., AV-rank = 1 and AV-rank = 10.
As shown in Figure~\ref{fig:maldeveloper}, under the threshold of AV-rank = 1, the vast majority of the developers (around 95\%) have published malicious apps, and about 55\% have \textit{Malware Rate} of 100\%, which means that they only release malware. Under the threshold of AV-rank = 10, there are still about 25\% of the developers with a \textit{Malware Rate} of 100\%.
Overall, for all the 228,443 apps we collected,  more than 77\% of them are flagged by at least one engine and roughly 55\% are flagged by at least 10 engines. \textit{This finding suggests that most of the apps released by these developers are malicious.}

\subsection{Origins and Targets}

\subsubsection{Developer Countries.}
\begin{figure}[!h]
\centering
\includegraphics[width=0.95\textwidth]{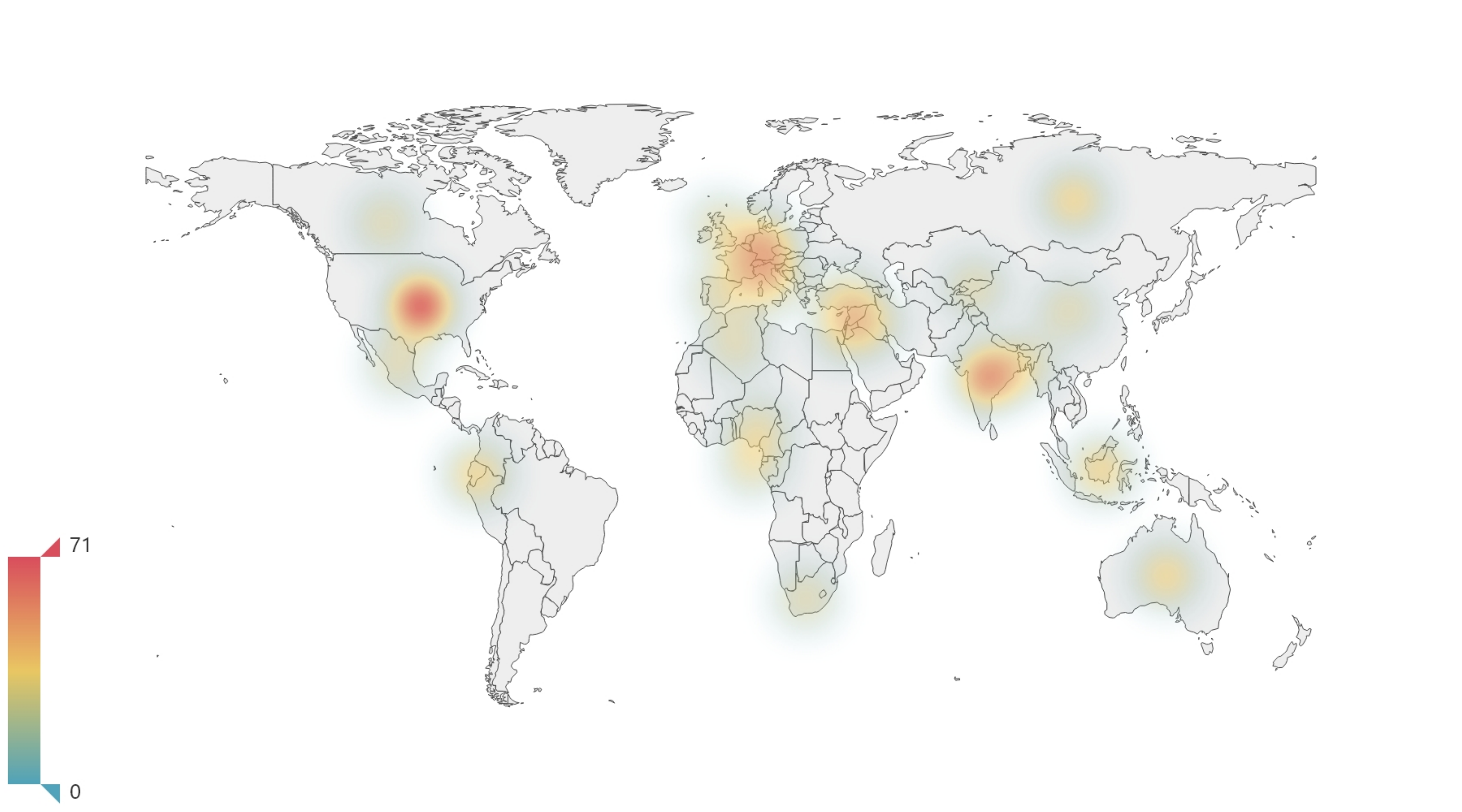}
\caption{The distribution of the countries of malicious developers. Most of them were claimed to be located in US.}
\label{fig:developer} 
\end{figure}

We are interested in learning about the countries of these malicious developers to investigate whether these malicious attacks are performed by developers in specific regions. However, it is non-trivial to know their real locations.
We can only extract their country information from the corresponding signature information. Note that, this country information might not be precise, as developers can intentionally modify it and provide a fake one, or just leave it empty. 
However, it is the only way for us to approximately investigate their countries. Finally, we have successfully identify the countries of 102 developers.
Figure~\ref{fig:developer} shows the distribution of the countries of malicious developers. Most of them (71 developers) claim to be located in the US, and the rest claim to be located in India, Germany, Australia, Indonesia, Russia, Italy, etc.

\begin{figure}[h]
\centering
\includegraphics[width=0.99\textwidth]{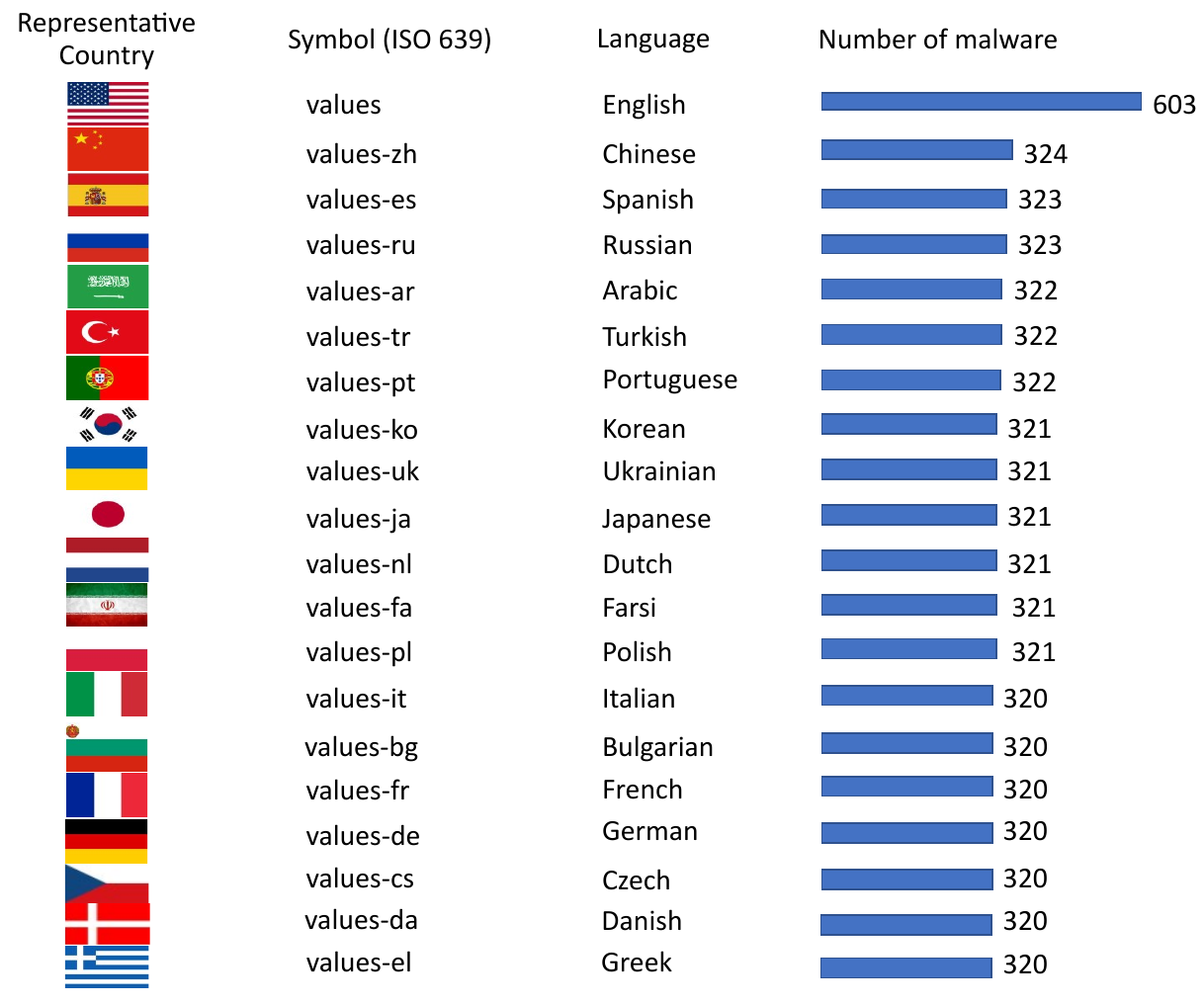}
\caption{Top 20 languages and their representative countries targeted by COVID-19 malware developers.}
\label{fig:language} 
\end{figure}

\subsubsection{Target Regions.}
We further want to know the target regions of these malicious apps, however, it is hard to know based only on the Android binary.
Here, we use an alternative approach.
The Android APK file stores some resource files under the \texttt{res/values} directory, such as \texttt{string.xml} and \texttt{arrays.xml}. 
After the app is launched, these resource files will be read and displayed on the UI. In order to display different languages texts on UIs in different countries or regions, Android app developers add different suffix strings to the \texttt{values} file names to distinguish languages they support and dynamically load these resource files when the app runs. These string names follow the \textit{ISO 639-3} encoding rule, which is an international language code standard that contains 136 two-letter codes for marking the world's major languages. These codes are used as shorthand for languages in many places, such as English is represented by \textit{en}, German is represented by \textit{de}, Chinese is represented by \textit{zh}.

We extract the names of all the \texttt{values} files under the \texttt{/res} folder, and compare these languages to check which countries or regions the apps can display. This data may indicate the countries and regions targeted by the malicious apps. Note that, developers not only use the region as a suffix, but also use the device screen resolution (such as values-hqpi, values-mdpi, etc.), and Android version (such as values-v19, values-v21, etc.) to display matching text information on different devices. Thus we filter out these kinds of files. Besides, this naming method also allows the area code to be added after the language to distinguish that multiple countries will use the same language, such as \texttt{values-pt-rBR}. 

Finally, we find that these malicious apps contain 87 different kinds of language resource files, of which 45\% contain only one language, 55\% contain at least 2 different languages. As shown in Figure~\ref{fig:language}, we list the top 20 languages supported by these malware and the representative regions. Actually, the 320 malware that contain the top-20 languages are the same. Starting in Android 7.0 (API level 24), Android provides enhanced support for multilingual users. It is easy for developers to include resources that can be specific to a particular culture and provide any resource type that is appropriate for the language and culture of the users when creating apps. 
Half of the malicious apps in our dataset contain more than 70 languages, making for a high degree of language overlap in these apps. This also explains the very close number of apps covering top-20 languages in Figure~\ref{fig:language}.
\texttt{English-speaking} countries are no doubt the primary target of the malware, roughly 99\% of apps support English. Besides, languages such as \texttt{Chinese}, \texttt{Spanish}, \texttt{Russian}, \texttt{Turkish} and \texttt{Arabic} are widely supported by these malicious apps.


\begin{framed}

\noindent \textbf{RQ \#$4$:} 
\textit{
Although 25\% of these malware creators are habitual developers that have been active for a long time, three quarters of the developers are newcomers in this coronavirus pandemic. 
Coronavirus is used as a lure to attack unsuspicious users.
Most of the apps released by these developers are malicious.
Based on the information collected, these developers are mainly located in the US, with rest of them are located in India, Turkey, etc. Besides English-speaking countries, China, Arabic countries, and Europe are also the main targets.
}
\end{framed}

\section{Discussion}
\label{sec:discussion}

\subsection{Implication}
This paper has several important observations that contribute to COVID-19 themed cybersecurity research. First, our findings suggest that malicious actors are quick at taking advantage of COVID-19 pandemic to perform cyber attacks. Although most of the malicious behaviors we identified (see Section~\ref{sec:behavior}) can be found in other non-COVID themed Android malware, a number of social engineering based techniques are used by COVID-19 themed malware to trick users.
Similar tricks can be easily adopted by them to other social event related attacks. Thus, 
users should pay special attention to the apps related to emerging social events. 
Second, our study in this paper suggests that some malicious apps have strong correlation that could be clustered into groups (see Section~\ref{sec:developers}) and roughly 25\% of the developers are habitual malicious developers that have been active for a long time, which could guide us to identify emerging malware and raise alarms at the early stage of their release. 

\subsection{Limitation}
We recognize that our study carries several limitations. First, our investigation is limited by apps we identified. Although we make efforts to build the seven most representative keywords to cover as many apps as possible through the four major channels, it is quite possible that some COVID-19 related apps are overlooked by us. Nevertheless, we believe our collection has covered most of the available COVID-19 themed apps. 
Second, the maliciousness of an app is determined by the scanned results of VirusTotal, which is the most common method but is not quite reliable because there is no standard on how to take advantage of the detection results to label malware and the results of VirusTotal may change over time. However, it must be acknowledged that it is unrealistic to manually detect whether each app is malicious or not. 
Third, some of the analysis relies on data that is not particularly accurate, such as extracting the developer's country from the signature information, which can be modified and controlled by the developer himself. But this is the only way we can think of to collect information about the developer's country.

\section{Related Work}

To the best of our knowledge, the coronavirus-themed mobile apps have not yet been systematically studied.
Nevertheless, various studies have explored the security and privacy aspects of mobile apps, as well as the studies of coronavirus pandemic from other domains. 

\subsection{Security Analysis of Mobile Apps}
A large mount of studies have analyzed mobile apps from security and privacy aspects, including malware detection~\cite{signature3,signature4,drebin}, permission and privacy analysis~\cite{wang2017understanding,wang2015reevaluating,wang2015using}, repackaging and fake app detection~\cite{wukong,hu2020mobile}, and identifying and analyzing third-party libraries~\cite{ma2016libradar,li2017libd,wang2017understanding}, etc.

Besides, some researchers in our community have analyzed specific types of mobile apps. For example, Hu et al.~\cite{hu2019dating} analyzed the ecosystem of fraudulent dating apps,
i.e., the sole purpose of these apps is to lure users into purchasing premium/VIP services to start conversations with other (likely fake female) accounts in the app. Ikram et al.~\cite{ikram2016analysis} measured 283 Android VPN apps to understand security and privacy issues.
Mobile health apps have been studied by previous work~\cite{sunyaev2015availability},~\cite{van2013mobile} and~\cite{grundy2016challenges}.
Our study suggests that the attackers are taking advantage of public events like COVID-19 to perform cyber-attacks. Although the malicious behaviors revealed in this paper are similar with other non-Coronavirus malware, they have adopted a number of social engineering techniques to deceive users, and similar techniques can be easily adopted to other social events. A number of existing tools and techniques can be adopted/integrated to analyze the issues in coronavirus-themed mobile apps. Thus, we decide to release the dataset to the community to boost the research on COVID-19 themed apps.
\textit{By the time of this writing, over 30 research institutes have requested our dataset for research purposes.}

\subsection{Coronavirus-related Studies}
Since its outbreak, COVID-19 has attracted great attentions from the research community.
A large number of studies were focused on the medical domain.
Many medical scientists have made outstanding contributions to the virus structure, pathological analysis, detection methods and treatment methods~\cite{wang2020clinical,chen2020sars,wrapp2020cryo,corman2020detection} of COVID-19.
Besides, a number of computer scientists have adopted machine learning techniques to identify and classify COVID-19 CT images. For example, Butt et al.~\cite{butt2020deep} designed multiple convolutional neural network (CNN) models to classify CT samples with COVID-19. Wang et al.~\cite{wang2020deep} used deep learning models to identify CT images of COVID-19 patients for fast judgment. 
In the field of social science, Kim~\cite{kim2020effects} collected the comments made by the Korean people on social media to analyze the negative emotions and social problems during the COVID-19 outbreak.
Lin et al.~\cite{lin2020google} used Google keyword search frequency to predict the speed of the spread of the COVID-19 outbreak in 21 countries/regions.
Schild et al.~\cite{schild2020go} collected comments from social media to analyze sinophobic behavior during the outbreak.
Malavolta~\cite{iivanoogithub} developed an automatic web scraper to crawl apps from Google Play and performed some basic analysis. However, only a few official apps were included and none of them is malware.
There is also a growing body of research on COVID-19 emerging in the software engineering community. For example, Neto et al.~\cite{neto2020deep} investigated the impact of COVID-19 on software projects and software development professionals by tapping into a library of software resources and survey research. As software developers migrate almost overnight to work from home, several studies have sought to understand developer productivity at technology companies~\cite{ford2020tale,Ralph_2020,bao2020does}.
Although a number of reports have revealed the existence of COVID-19 themed Android malware, to the best of our knowledge, our study is the first to characterize them in a systematic way.

\section{Conclusion}

In this paper, we present the first measurement study of COVID-19 themed Android malware. We first make efforts to create and maintain a repository of COVID-19 themed apps, by collecting samples from a number of sources, including app markets, a well-known app repository, the COVID-19 related domains and security threat intelligence reports. We then present comprehensive analysis of these apps from the perspectives of trends and statistics, distribution and installation, malicious behaviors, and the attackers behind them. 
Our observations suggest that malicious actors are quick at taking advantage of COVID-19 pandemic to perform cyber attacks.
Our study can help boost the research on social event based cyber security threats.


\bibliographystyle{spmpsci}
\bibliography{cite}

\begin{thebibliography}{10}
\providecommand{\url}[1]{{#1}}
\providecommand{\urlprefix}{URL }
\expandafter\ifx\csname urlstyle\endcsname\relax
  \providecommand{\doi}[1]{DOI~\discretionary{}{}{}#1}\else
  \providecommand{\doi}{DOI~\discretionary{}{}{}\begingroup
  \urlstyle{rm}\Url}\fi

\bibitem{AlienVault}
{AlienVault}.
\newblock https://otx.alienvault.com/

\bibitem{MalwareFamily}
{Malware Family}.
\newblock https://www.sciencedirect.com/topics/computer-science/malware-family

\bibitem{threat-intelligence-platforms}
{Threat Intelligence Platform}.
\newblock
  https://www.esecurityplanet.com/products/threat-intelligence-platforms/

\bibitem{apkid}
{APKiD}.
\newblock https://github.com/rednaga/APKiD (2020)

\bibitem{ApkProtect}
{ApkProtect}.
\newblock https://apkprotect.baidu.com/ (2020)

\bibitem{Apkpure}
Apkpure.
\newblock https://apkpure.com (2020)

\bibitem{appchina}
{AppChina}.
\newblock http://www.appchina.com (2020)

\bibitem{App-SocialEngineering-2}
{Computer and Mobile Based Social Engineering}.
\newblock https://www.greycampus.com/opencampus
  /ethical-hacking/computer-and-mobile-based-social-engineering (2020)

\bibitem{paloalto}
{COVID-19: Cloud Threat Landscape}.
\newblock https://unit42.paloaltonetworks.com/covid-19-cloud-threat-landscape/
  (2020)

\bibitem{NCSC}
{COVID-19 Exploited by Malicious Cyber Actors}.
\newblock https://www.us-cert.gov/ncas/alerts/aa20-099a (2020)

\bibitem{androidreport3}
{COVID-19 goes mobile: Coronavirus malicious applications discovered}.
\newblock
  https://research.checkpoint.com/2020/covid-19-goes-mobile-coronavirus-malicious-applications-discovered/
  (2020)

\bibitem{androidreport4}
{COVID-19-Themed Malware Goes Mobile}.
\newblock
  https://www.bankinfosecurity.com/covid-19-themed-malware-goes-mobile-a-13981
  (2020)

\bibitem{mcafeemalware}
{COVID-19 – Malware Makes Hay During a Pandemic}.
\newblock
  https://www.mcafee.com/blogs/other-blogs/mcafee-labs/covid-19-malware-makes-hay-during-a-pandemic/
  (2020)

\bibitem{trendmicronews}
{Developing Story: COVID-19 Used in Malicious Campaigns}.
\newblock
  https://www.trendmicro.com/vinfo/us/security/news/cybercrime-and-digital-threats/coronavirus-used-in-spam-malware-file-names-and-malicious-domains
  (2020)

\bibitem{App-SocialEngineering-1}
{Fake Netflix Android app is social engineering scam}.
\newblock
  https://www.csoonline.com/article/2129805/fake-netflix-android-app-is-social-engineering-scam.html
  (2020)

\bibitem{androidreport2}
{Findings on COVID-19 and online security threats}.
\newblock
  https://www.blog.google/technology/safety-security/threat-analysis-group/findings-covid-19-and-online-security-threats/
  (2020)

\bibitem{bankinfosecurity}
{Fresh COVID-19 Phishing Scams Try to Spread Malware: Report}.
\newblock
  https://www.bankinfosecurity.com/fresh-covid-19-phishing-scams-try-to-spread-malware-report-a-14131
  (2020)

\bibitem{Googleplay}
{Google Play}.
\newblock https://play.google.com (2020)

\bibitem{Huawei}
{Huawei Market}.
\newblock http://app.hicloud.com (2020)

\bibitem{Koodous}
Koodous.
\newblock https://koodous.com/ (2020)

\bibitem{MalwareClassification}
{Malware Classifications}.
\newblock
  https://www.kaspersky.com/resource-center/threats/malware-classifications
  (2020)

\bibitem{microsoftcategory}
Malware names.
\newblock
  https://docs.microsoft.com/en-us/windows/security/threat-protection/intelligence/malware-naming
  (2020)

\bibitem{MyApp}
{MyApp Market}.
\newblock https://android.myapp.com (2020)

\bibitem{MYSTORY}
{MYSTORY}.
\newblock
  https://yourstory.com/mystory/smartphones-prime-targets-cybercriminals (2020)

\bibitem{androidreport1}
{New Android Coronavirus Malware Threat Exposed: Here’s What You Must Not
  Do}.
\newblock
  https://www.forbes.com/sites/zakdoffman/2020/04/09/why-android-users-must-now-dodge-this-simple-15-minute-coronavirus-malware-threat/\#6b020abc4c1d
  (2020)

\bibitem{scienceApp}
{Show evidence that apps for COVID-19 contact-tracing are secure and
  effective}.
\newblock https://www.nature.com/articles/d41586-020-01264-1 (2020)

\bibitem{SocialEngineering-1}
{Social Engineering}.
\newblock
  https://www.imperva.com/learn/application-security/social-engineering-attack/
  (2020)

\bibitem{SocialEngineering-2}
{Top 5 Social Engineering Techniques and How to Prevent Them}.
\newblock https://www.exabeam.com/information-security/social-engineering/
  (2020)

\bibitem{uptodown}
Uptodown.
\newblock https://en.uptodown.com (2020)

\bibitem{urlscan}
Urlscan.
\newblock https://urlscan.io (2020)

\bibitem{fireeye}
{Vietnamese Threat Actors APT32 Targeting Wuhan Government and Chinese Ministry
  of Emergency Management in Latest Example of COVID-19 Related Espionage}.
\newblock
  https://www.fireeye.com/blog/threat-research/2020/04/apt32-targeting-chinese-government-in-covid-19-related-espionage.html
  (2020)

\bibitem{VirusTotal}
{VirusTotal}.
\newblock https://www.virustotal.com/ (2020)

\bibitem{iivanoogithub}
{Web scraper and analyzer of COVID-related Android apps}.
\newblock https://github.com/iivanoo/covid-apps-observer (2020)

\bibitem{drebin}
Arp, D., Spreitzenbarth, M., Hubner, M., Gascon, H., Rieck, K., Siemens, C.:
  Drebin: Effective and explainable detection of android malware in your
  pocket.
\newblock In: Ndss, vol.~14, pp. 23--26 (2014)

\bibitem{arzt2014flowdroid}
Arzt, S., Rasthofer, S., Fritz, C., Bodden, E., Bartel, A., Klein, J.,
  Le~Traon, Y., Octeau, D., McDaniel, P.: Flowdroid: Precise context, flow,
  field, object-sensitive and lifecycle-aware taint analysis for android apps.
\newblock Acm Sigplan Notices \textbf{49}(6), 259--269 (2014)

\bibitem{bao2020does}
Bao, L., Li, T., Xia, X., Zhu, K., Li, H., Yang, X.: How does working from home
  affect developer productivity? -- a case study of baidu during covid-19
  pandemic (2020)

\bibitem{butt2020deep}
Butt, C., Gill, J., Chun, D., Babu, B.A.: Deep learning system to screen
  coronavirus disease 2019 pneumonia.
\newblock Applied Intelligence p.~1 (2020)

\bibitem{chen2020sars}
Chen, Y., Li, L.: Sars-cov-2: virus dynamics and host response.
\newblock The Lancet Infectious Diseases \textbf{20}(5), 515--516 (2020)

\bibitem{corman2020detection}
Corman, V.M., Landt, O., Kaiser, M., Molenkamp, R., Meijer, A., Chu, D.K.,
  Bleicker, T., Br{\"u}nink, S., Schneider, J., Schmidt, M.L., et~al.:
  Detection of 2019 novel coronavirus (2019-ncov) by real-time rt-pcr.
\newblock Eurosurveillance \textbf{25}(3), 2000045 (2020)

\bibitem{DAVIS2018224}
Davis, B., Hasson, U.: Predictability of what or where reduces brain activity,
  but a bottleneck occurs when both are predictable.
\newblock NeuroImage \textbf{167}, 224 -- 236 (2018).
\newblock \doi{https://doi.org/10.1016/j.neuroimage.2016.06.001}.
\newblock
  \urlprefix\url{http://www.sciencedirect.com/science/article/pii/S1053811916302014}

\bibitem{https://doi.org/10.1002/jeab.176}
Diamond, R.F.L., Stoinski, T.S., Mickelberg, J.L., Basile, B.M., Gazes, R.P.,
  Templer, V.L., Hampton, R.R.: Similar stimulus features control visual
  classification in orangutans and rhesus monkeys.
\newblock Journal of the Experimental Analysis of Behavior \textbf{105}(1),
  100--110.
\newblock \doi{https://doi.org/10.1002/jeab.176}.
\newblock
  \urlprefix\url{https://onlinelibrary.wiley.com/doi/abs/10.1002/jeab.176}

\bibitem{Farooqi2020UnderstandingIM}
Farooqi, S., Feal, {\'A}., Lauinger, T., McCoy, D., Shafiq, Z.,
  Vallina-Rodriguez, N.: Understanding incentivized mobile app installs on
  google play store.
\newblock Proceedings of the ACM Internet Measurement Conference  (2020)

\bibitem{signature4}
Feng, Y., Anand, S., Dillig, I., Aiken, A.: Apposcopy: Semantics-based
  detection of android malware through static analysis.
\newblock In: Proceedings of the 22nd ACM SIGSOFT International Symposium on
  Foundations of Software Engineering, pp. 576--587 (2014)

\bibitem{ford2020tale}
Ford, D., Storey, M.A., Zimmermann, T., Bird, C., Jaffe, S., Maddila, C.,
  Butler, J.L., Houck, B., Nagappan, N.: A tale of two cities: Software
  developers working from home during the covid-19 pandemic (2020)

\bibitem{EvaluationOfResourceBasedAppRepackagingDetection_Gadyatskaya2016}
Gadyatskaya, O., Lezza, A.L., Zhauniarovich, Y.: {Evaluation of Resource-based
  App Repackaging Detection in Android}.
\newblock In: Proceedings of the 21st Nordic Conference on Secure IT Systems,
  NordSec 2016, pp. 135--151 (2016)

\bibitem{6107902}
{Gennari}, J., {French}, D.: Defining malware families based on analyst
  insights.
\newblock In: 2011 IEEE International Conference on Technologies for Homeland
  Security (HST), pp. 396--401 (2011).
\newblock \doi{10.1109/THS.2011.6107902}

\bibitem{grundy2016challenges}
Grundy, Q.H., Wang, Z., Bero, L.A.: Challenges in assessing mobile health app
  quality: a systematic review of prevalent and innovative methods.
\newblock American journal of preventive medicine \textbf{51}(6), 1051--1059
  (2016)

\bibitem{hu2020mobile}
Hu, Y., Wang, H., He, R., Li, L., Tyson, G., Castro, I., Guo, Y., Wu, L., Xu,
  G.: Mobile app squatting.
\newblock In: Proceedings of The Web Conference 2020, pp. 1727--1738 (2020)

\bibitem{hu2019dating}
Hu, Y., Wang, H., Zhou, Y., Guo, Y., Li, L., Luo, B., Xu, F.: Dating with
  scambots: understanding the ecosystem of fraudulent dating applications.
\newblock IEEE Transactions on Dependable and Secure Computing  (2019)

\bibitem{ikram2016analysis}
Ikram, M., Vallina-Rodriguez, N., Seneviratne, S., Kaafar, M.A., Paxson, V.: An
  analysis of the privacy and security risks of android vpn permission-enabled
  apps.
\newblock In: Proceedings of the 2016 Internet Measurement Conference, pp.
  349--364 (2016)

\bibitem{IYENGAR2020733}
Iyengar, K., Upadhyaya, G.K., Vaishya, R., Jain, V.: Covid-19 and applications
  of smartphone technology in the current pandemic.
\newblock Diabetes \& Metabolic Syndrome: Clinical Research \& Reviews
  \textbf{14}(5), 733 -- 737 (2020).
\newblock \doi{https://doi.org/10.1016/j.dsx.2020.05.033}

\bibitem{kim2020effects}
Kim, B.: Effects of social grooming on incivility in covid-19.
\newblock Cyberpsychology, Behavior, and Social Networking  (2020)

\bibitem{KWON2014137}
Kwon, T., Na, S.: Tinylock: Affordable defense against smudge attacks on
  smartphone pattern lock systems.
\newblock Computers \& Security \textbf{42}, 137 -- 150 (2014).
\newblock \doi{https://doi.org/10.1016/j.cose.2013.12.001}.
\newblock
  \urlprefix\url{http://www.sciencedirect.com/science/article/pii/S0167404813001697}

\bibitem{kywe2014detecting}
Kywe, S.M., Li, Y., Deng, R.H., Hong, J.: Detecting camouflaged applications on
  mobile application markets.
\newblock In: International Conference on Information Security and Cryptology,
  pp. 241--254. Springer (2014)

\bibitem{piggybacking}
Li, L., Li, D., Bissyand{\'e}, T.F., Klein, J., Le~Traon, Y., Lo, D.,
  Cavallaro, L.: Understanding android app piggybacking: A systematic study of
  malicious code grafting.
\newblock IEEE Transactions on Information Forensics and Security
  \textbf{12}(6), 1269--1284 (2017)

\bibitem{li2017libd}
Li, M., Wang, W., Wang, P., Wang, S., Wu, D., Liu, J., Xue, R., Huo, W.: Libd:
  scalable and precise third-party library detection in android markets.
\newblock In: 2017 IEEE/ACM 39th International Conference on Software
  Engineering (ICSE), pp. 335--346. IEEE (2017)

\bibitem{li2017droidbot}
Li, Y., Yang, Z., Guo, Y., Chen, X.: Droidbot: a lightweight ui-guided test
  input generator for android.
\newblock In: 2017 IEEE/ACM 39th International Conference on Software
  Engineering Companion (ICSE-C), pp. 23--26. IEEE (2017)

\bibitem{lin2020google}
Lin, Y.H., Liu, C.H., Chiu, Y.C.: Google searches for the keywords of “wash
  hands” predict the speed of national spread of covid-19 outbreak among 21
  countries.
\newblock Brain, Behavior, and Immunity  (2020)

\bibitem{MadDroid}
Liu, T., Wang, H., Li, L., Luo, X., Dong, F., Guo, Y., Wang, L., Bissyande,
  T.F., Klein, J.: Maddroid: Characterising and detecting devious ad content
  for android apps.
\newblock In: Proceedings of the Web Conference 2020 (WWW'20) (2020)

\bibitem{ma2016libradar}
Ma, Z., Wang, H., Guo, Y., Chen, X.: Libradar: fast and accurate detection of
  third-party libraries in android apps.
\newblock In: Proceedings of the 38th international conference on software
  engineering companion, pp. 653--656 (2016)

\bibitem{neto2020deep}
da~Mota Silveira~Neto, P.A., Mannan, U.A., de~Almeida, E.S., Nagappan, N., Lo,
  D., Kochhar, P.S., Gao, C., Ahmed, I.: A deep dive on the impact of covid-19
  in software development (2020)

\bibitem{Ralph_2020}
Ralph, P., Baltes, S., Adisaputri, G., Torkar, R., Kovalenko, V., Kalinowski,
  M., Novielli, N., Yoo, S., Devroey, X., Tan, X., et~al.: Pandemic
  programming.
\newblock Empirical Software Engineering  (2020).
\newblock \doi{10.1007/s10664-020-09875-y}.
\newblock \urlprefix\url{http://dx.doi.org/10.1007/s10664-020-09875-y}

\bibitem{rasthofer2016harvesting}
Rasthofer, S., Arzt, S., Miltenberger, M., Bodden, E.: Harvesting runtime
  values in android applications that feature anti-analysis techniques.
\newblock In: NDSS (2016)

\bibitem{schild2020go}
Schild, L., Ling, C., Blackburn, J., Stringhini, G., Zhang, Y., Zannettou, S.:
  " go eat a bat, chang!": An early look on the emergence of sinophobic
  behavior on web communities in the face of covid-19.
\newblock arXiv preprint arXiv:2004.04046  (2020)

\bibitem{sebastian2016avclass}
Sebasti{\'a}n, M., Rivera, R., Kotzias, P., Caballero, J.: Avclass: A tool for
  massive malware labeling.
\newblock In: International Symposium on Research in Attacks, Intrusions, and
  Defenses, pp. 230--253. Springer (2016)

\bibitem{sunyaev2015availability}
Sunyaev, A., Dehling, T., Taylor, P.L., Mandl, K.D.: Availability and quality
  of mobile health app privacy policies.
\newblock Journal of the American Medical Informatics Association
  \textbf{22}(e1), e28--e33 (2015)

\bibitem{van2013mobile}
van Velsen, L., Beaujean, D.J., van Gemert-Pijnen, J.E.: Why mobile health app
  overload drives us crazy, and how to restore the sanity.
\newblock BMC medical informatics and decision making \textbf{13}(1), 23 (2013)

\bibitem{wang2020clinical}
Wang, D., Hu, B., Hu, C., Zhu, F., Liu, X., Zhang, J., Wang, B., Xiang, H.,
  Cheng, Z., Xiong, Y., et~al.: Clinical characteristics of 138 hospitalized
  patients with 2019 novel coronavirus--infected pneumonia in wuhan, china.
\newblock Jama \textbf{323}(11), 1061--1069 (2020)

\bibitem{wukong}
Wang, H., Guo, Y., Ma, Z., Chen, X.: Wukong: a scalable and accurate two-phase
  approach to android app clone detection.
\newblock In: Proceedings of the 2015 International Symposium on Software
  Testing and Analysis, pp. 71--82. ACM (2015)

\bibitem{wang2015reevaluating}
Wang, H., Guo, Y., Tang, Z., Bai, G., Chen, X.: Reevaluating android permission
  gaps with static and dynamic analysis.
\newblock In: 2015 IEEE Global Communications Conference (GLOBECOM), pp. 1--6.
  IEEE (2015)

\bibitem{wang2015using}
Wang, H., Hong, J., Guo, Y.: Using text mining to infer the purpose of
  permission use in mobile apps.
\newblock In: Proceedings of the 2015 ACM International Joint Conference on
  Pervasive and Ubiquitous Computing, pp. 1107--1118 (2015)

\bibitem{wang2019understanding}
Wang, H., Li, H., Guo, Y.: Understanding the evolution of mobile app
  ecosystems: A longitudinal measurement study of google play.
\newblock In: The World Wide Web Conference, pp. 1988--1999 (2019)

\bibitem{wang2017understanding}
Wang, H., Li, Y., Guo, Y., Agarwal, Y., Hong, J.I.: Understanding the purpose
  of permission use in mobile apps.
\newblock ACM Transactions on Information Systems (TOIS) \textbf{35}(4), 1--40
  (2017)

\bibitem{wang2019characterizing}
Wang, H., Liu, H., Xiao, X., Meng, G., Guo, Y.: Characterizing android app
  signing issues.
\newblock In: 2019 34th IEEE/ACM International Conference on Automated Software
  Engineering (ASE), pp. 280--292. IEEE (2019)

\bibitem{wang2018beyond}
Wang, H., Liu, Z., Liang, J., Vallina-Rodriguez, N., Guo, Y., Li, L., Tapiador,
  J., Cao, J., Xu, G.: Beyond google play: A large-scale comparative study of
  chinese android app markets.
\newblock In: Proceedings of the Internet Measurement Conference 2018, pp.
  293--307 (2018)

\bibitem{wang2020deep}
Wang, S., Kang, B., Ma, J., Zeng, X., Xiao, M., Guo, J., Cai, M., Yang, J., Li,
  Y., Meng, X., et~al.: A deep learning algorithm using ct images to screen for
  corona virus disease (covid-19).
\newblock MedRxiv  (2020)

\bibitem{wang2017droid}
Wang, X., Zhu, S., Zhou, D., Yang, Y.: Droid-antirm: Taming control flow
  anti-analysis to support automated dynamic analysis of android malware.
\newblock In: Proceedings of the 33rd Annual Computer Security Applications
  Conference, pp. 350--361 (2017)

\bibitem{wrapp2020cryo}
Wrapp, D., Wang, N., Corbett, K.S., Goldsmith, J.A., Hsieh, C.L., Abiona, O.,
  Graham, B.S., McLellan, J.S.: Cryo-em structure of the 2019-ncov spike in the
  prefusion conformation.
\newblock Science \textbf{367}(6483), 1260--1263 (2020)

\bibitem{signature3}
Zhang, M., Duan, Y., Yin, H., Zhao, Z.: Semantics-aware android malware
  classification using weighted contextual api dependency graphs.
\newblock In: Proceedings of the 2014 ACM SIGSAC conference on computer and
  communications security, pp. 1105--1116 (2014)

\bibitem{zhou2012detecting}
Zhou, W., Zhou, Y., Jiang, X., Ning, P.: Detecting repackaged smartphone
  applications in third-party android marketplaces.
\newblock In: Proceedings of the second ACM conference on Data and Application
  Security and Privacy, pp. 317--326 (2012)

\bibitem{zhou2012dissecting}
Zhou, Y., Jiang, X.: Dissecting android malware: Characterization and
  evolution.
\newblock In: 2012 IEEE symposium on security and privacy, pp. 95--109. IEEE
  (2012)

\end{thebibliography}

\end{document}